\pdfoutput=1
\RequirePackage{ifpdf}
\ifpdf 
\documentclass[pdftex]{sigma}
\else
\documentclass{sigma}
\fi

\numberwithin{equation}{section}
\newtheorem{thm}{Theorem}[section]

{ \theoremstyle{definition}
\newtheorem{Def}[thm]{Definition}
}

\newcommand{\B}{\mathbf}

\newcommand{\ee}{\textup{e}}
\newcommand{\dd}{\textup{d}}
\newcommand{\ii}{\textup{i}}
\newcommand{\la}{\lambda}
\newcommand{\ord}{\mathrm{O}}
\renewcommand{\Im}{\operatorname{Im}}
\renewcommand{\Re}{\operatorname{Re}}

\begin{document}
\allowdisplaybreaks

\newcommand{\arXivNumber}{1802.01622}

\renewcommand{\PaperNumber}{082}

\FirstPageHeading

\ShortArticleName{A Matrix Baker--Akhiezer Function Associated with the Maxwell--Bloch Equations}

\ArticleName{A Matrix Baker--Akhiezer Function\\ Associated with the Maxwell--Bloch Equations\\ and their Finite-Gap Solutions}

\Author{Vladimir P.~KOTLYAROV}

\AuthorNameForHeading{V.P.~Kotlyarov}

\Address{B.~Verkin Institute for Low Temperature Physics and Engineering,\\ 47~Lenin Ave., 61103 Kharkiv, Ukraine}
\Email{\href{mailto:kotlyarov@ilt.kharkov.ua}{kotlyarov@ilt.kharkov.ua}}

\ArticleDates{Received February 05, 2018, in final form August 02, 2018; Published online August 10, 2018}

\Abstract{The Baker--Akhiezer (BA) function theory was successfully developed in the mid 1970s. This theory brought very interesting and important results in the spectral theory of almost periodic operators and theory of completely integrable nonlinear equations such as Korteweg--de Vries equation, nonlinear Schr\"odinger equation, sine-Gordon equation, Kadomtsev--Petviashvili equation. Subsequently the theory was reproduced for the Ablowitz--Kaup--Newell--Segur (AKNS) hierarchies. However, extensions of the Baker--Akhiezer function for the Maxwell--Bloch (MB) system or for the Karpman--Kaup equations, which contain prescribed weight functions characterizing inhomogeneous broadening of the main frequency, are unknown. The main goal of the paper is to give a such of extension associated with the Maxwell--Bloch equations. Using different Riemann--Hilbert problems posed on the complex plane with a finite number of cuts we propose such a matrix function that has unit determinant and takes an explicit form through Cauchy integrals, hyperelliptic integrals and theta functions. The matrix BA function solves the AKNS equations (the Lax pair for MB system) and generates a quasi-periodic finite-gap solution to the Maxwell--Bloch equations. The suggested function will be useful in the study of the long time asymptotic behavior of solutions of different initial-boundary value problems for the MB equations using the Deift--Zhou method of steepest descent and for an investigation of rogue waves of the Maxwell--Bloch equations.}

\Keywords{Baker--Akhiezer function; Maxwell--Bloch equations; matrix Riemann--Hilbert problems}
\Classification{34L25; 34M50; 35F31; 35Q15; 35Q51}

\section{Introduction}

We consider the Maxwell--Bloch (MB) equations written in the form
\begin{gather} \label{MB1}
{\mathcal E}_t+{\mathcal E}_x =\langle\rho\rangle, \qquad \langle\rho\rangle=\Omega\int_{-\infty}^\infty n(\lambda)\rho(t,x,\lambda)\dd\lambda,\\
\rho_t+2\ii\lambda\rho={\mathcal N}{\mathcal E},\label{MB2}\\
{\mathcal N}_t =-\frac{1}{2}({\mathcal E}^*\rho+{\mathcal E}\rho^*).\label{MB3}
\end{gather}
Here, ${\mathcal E}={\mathcal E}(t,x)$ is a complex-valued function of the time $t$ and the coordinate $x$, and $\rho =\rho(t, x, \lambda)$ and ${\mathcal N} ={\mathcal N}(t, x, \lambda)$ are complex-valued and real functions of~$ t$, $x$, and the additional parameter $\lambda$. Subindices refer to partial derivatives in~$t$ and~$x$, and~$*$ means a~complex conjugation.

Equations (\ref{MB1})--(\ref{MB3}) are used in many physical models which deal with a classical electromagnetic field that interacts resonantly with quantum two-level objects~-- two-level atoms, which have only two energy position: upper and lower level. In particular, there are models of the self-induced transparency \cite{AKN,AS}, and two-level laser amplifier \cite{M,MN}. For these models ${{\mathcal E}=\mathcal E}(t,x)$ is the complex valued envelope of an electromagnetic wave of fixed polarization, so that the field in the resonant medium is
\begin{gather*}
{\bf E}(t,x)={\mathcal E}(t,x)e^{\ii\Omega(x-t)}+{\mathcal E}^*(t,x)e^{-\ii\Omega(x-t)}.
\end{gather*}
${\cal N}(t,x,\lambda)$ and ${\rho}(t,x,\lambda)$ are entries of the density matrix $F(t,x,\lambda)=\left(\begin{smallmatrix}{\cal N}(t,x,\lambda)&\rho(t,x,\lambda)\\
 \rho^*(t,x,\lambda)&-{\cal N}(t,x,\lambda)\end{smallmatrix}\right)$. It describes the atomic subsystem. The parameter $\la$ is the deviation of transition frequency of given two-level atom from its mean frequency~$\Omega$. The angular brackets in~\eqref{MB1} mean averaging with given weight function $n(\lambda)>0$, such that
\begin{gather}\label{n}
\int_{-\infty}^\infty n(\lambda)\dd\lambda=1.
\end{gather}
The weight function $n(\lambda)$ characterizes inhomogeneous broadening. From \eqref{MB2} and \eqref{MB3} it follows that
\begin{gather*}
\frac{\partial}{\partial t}\big({\mathcal N}^2(t,x,\lambda)+|\rho(t,x,\lambda)|^2\big)=0.
\end{gather*}
We interest in solutions where initial data are subjected to the condition
\begin{gather*}
{\mathcal N}^2(0,x,\lambda)+|\rho(0,x,\lambda)|^2\equiv 1.
\end{gather*}
Then
\begin{gather*}
{\mathcal N}^2(t,x,\lambda)+|\rho(t,x,\lambda)|^2\equiv 1
\end{gather*}
for all~$t$, which reflects the conservation of probability: the total probability that an atom can be found in the upper or lower level equals~1. We also put $\Omega=1$ in \eqref{MB1}. For a given (at the initial time) polarization, the population is determined to within a sign
\begin{gather*}
{\mathcal N}(0,x,\lambda)=\pm\sqrt{1-|\rho(0,x,\lambda)|^2}.
\end{gather*}
If ${\mathcal N}(0,x,\lambda)>0$, then an unstable medium is considered (the so-called two-level laser amplifier). If ${\mathcal N}(0,x,\lambda)<0$, then a stable medium is considered (the so-called attenuator).

The Maxwell--Bloch equations became well-known in soliton theory after Lamb \cite{L1,L2,L3,L4}. Ablowitz, Kaup and Newell have firstly applied the inverse scattering transform to the Maxwell--Bloch equations in \cite{AKN}. In some sense general solutions to the MB equations and their classifying were done by Gabitov, Zakharov and Mikhailov in~\cite{GZM}. Some asymptotic results for the MB equations were obtained by Manakov in~\cite{M} and, in a collaboration with Novokshenov, in~\cite{MN}. Elliptic periodic waves in the theory of self-induced transparency were constructed by Kamchatnov in~\cite{Kam95}. We cite here only a small number of pioneering papers relating to the Maxwell--Bloch equations. Some reviews on an application of inverse scattering transform to the MB equations can be found in \cite{AKN,AS,GZM,Kis}, and for the reduced Maxwell--Bloch equations in \cite{HC, WWG}.

A Lax pair for the Maxwell--Bloch system was first found in \cite{AKN} by using results of \cite{L1,L2,L3,L4} (see also \cite{AS, GZM}). It was shown that \eqref{MB1}--\eqref{MB3} are the compatibility condition of an overdetermined linear system, known as the Ablowitz--Kaup--Newell--Segur (AKNS) equations
\begin{gather}
w_t+\ii\la\sigma_3 w=-H(t,x)w, \label{AS1}\\
w_x-\ii\la\sigma_3 +\ii G(t,x,\la)w=H(t,x)w, \label{AS2}
\end{gather}
where
\begin{gather*}
\sigma_3 =\begin{pmatrix}1&0\\0&-1\end{pmatrix},\qquad
H(t,x)=\frac{1}{2}
\begin{pmatrix}0&{\mathcal E}(t,x)\\
-{\mathcal E}^*(t,x)&0\end{pmatrix}, \\
G(t,x,\lambda)= {\rm p.v.}\frac{1}{4}\int_{-\infty}^\infty \frac{F(t,x,s)n(s)}{s-\lambda}\dd s.
\end{gather*}

The symbol $\rm{p.v.}$ denotes the principal value integral. Differential equations \eqref{AS1} and \eqref{AS2} are compatible if and only if ${\mathcal E}(t,x)$, $\rho(t,x,\lambda)$ and ${\mathcal N}(t,x,\lambda)$ satisfy equations \eqref{MB1}--\eqref{MB3} (see, for example,~\cite{AS}). As shown in \cite{L4}, $\rho(t,x,\lambda)$ and ${\mathcal N}(t,x,\lambda)$ are related to the fundamental matrix of \eqref{AS1}. Indeed, let $\varPhi(t,x,\lambda)$ be a solution of \eqref{AS1} such that $\det\varPhi(t,x,\lambda)\equiv1$ and $\varPhi^\dag$ be the Hermitian-conjugated to $\varPhi$. Then $F(t,x,\lambda)=\varPhi(t,x,\lambda)\sigma_3\varPhi^\dag(t,x,\lambda)$ satisfies equation
\begin{gather*}
F_t+[\ii\la\sigma_3+H, F]=0.
\end{gather*}
It is a matrix form of the equations \eqref{MB2} and \eqref{MB3}.

In some cases it is convenient \cite{K} to use equations
\begin{gather}\label{pmxeq}
w_x-\ii\lambda\sigma_3 w +\ii G_\pm(t,x,\lambda) w=H(t,x) w,
\end{gather}
where
\begin{gather*}
G_\pm(t,x,\lambda)=\frac{1}{4}\int_{-\infty}^\infty
\frac{F(t,x,s) n(s)}{s-\lambda\mp\ii 0}\dd s={\rm p.v.}\frac{1}{4}\int_{-\infty}^\infty
 \frac{F(t,x,s)n(s)}{s-\la }\dd s\pm\frac{\pi\ii}{4}F(t,x,\la)n(\la).
\end{gather*}
Thus there are two Lax pairs ($t$- and $x^+$-equations and $t$- and $x^-$-equations) for the MB equations. Equations (\ref{AS1}) and (\ref{pmxeq}) (as well as (\ref{AS1}) and (\ref{AS2})) are compatible if and only if ${\mathcal E}(t,x)$, $\rho(t,x,\lambda)$ and ${\mathcal N}(t,x,\lambda)$ satisfy equations (\ref{MB1})--(\ref{MB3}).

The main goal of the paper is to give a construction of the Baker--Akhiezer function $\Psi(t,x,z)$ associated with the Maxwell--Bloch equations. Using different Riemann--Hilbert problems posed on the complex plane with a finite number of cuts we propose such a matrix function $\Psi(t,x,z)$ that has unit determinant and takes an explicit form through theta functions and Cauchy integrals. The construction proceeds also from the requirement that $\Psi(t,x,z)$ must satisfy the following system of linear equations
\begin{gather} \label{zt}
w_t+\ii z\sigma_3 w=-H(t,x) w,\\ \label{zx}
w_x-\ii z\sigma_3 w +\ii G(t,x,z) w=H(t,x) w,
\end{gather}
which depend on $z\in\mathbb{C}\setminus\Sigma$ where $\Sigma$ is a contour containing (as a part) the real axis $\mathbb{R}$ of the complex plane, and
\begin{gather*}
G(t,x,z)=\frac{1}{4}\int_{-\infty}^\infty \frac{F(t,x,s) n(s)}{s-z}\dd s, \qquad \Im z\neq0.
\end{gather*}
Symmetries of $F$, $G$, $H$ and equations \eqref{zt}, \eqref{zx} provide the following symmetry of $\Psi$
\begin{gather}\label{symPsi}
\Psi(t,x,z)=\sigma_2\Psi^*(t,x,z^*)\sigma_2, \qquad \sigma_2=\begin{pmatrix}
0 & -\ii \\
\ii & 0 \end{pmatrix}.
\end{gather}
As a result of our construction we obtain also a solution to the Maxwell--Bloch equations \eqref{MB1}--\eqref{MB3}. This solution is an analog of finite-gap solutions of soliton equations.

The Baker--Akhiezer function theory, as an analogue of the Floquet theory for ODE's with periodic coefficients, was successfully developed many years ago, in the mid 1970s. This theory brought very interesting and important results in the spectral theory of almost periodic operators and theory of completely integrable nonlinear equations such as Korteweg--de Vries equation, nonlinear Schr\"odinger equation, sine-Gordon equation, Kadomtsev--Petviashvili equation (see, e.g., \cite{BBEIM,D1,DKN,DMN,IM1,IM2,IM3,IM4,Kri1,Kri2,M1,N74}). Subsequently the theory was reproduced for the Ablowitz--Kaup--Newell--Segur (AKNS) hierarchies. However, extensions of the Baker--Akhiezer function for the Maxwell--Bloch system or for the Karpman--Kaup equations \cite{HL, Karp}, which contain prescribed weight functions characterizing inhomogeneous broadening of the main frequency, are unknown. The main goal of the paper is to give a such of extension associated with the Maxwell--Bloch equations. One more goal is applications in asymptotic analysis. The presence of inhomogeneous broadening~$n(\lambda)$ leads to noticeable complications in the Deift--Zhou method of steepest descent \cite{DIZ93, DIZ97,DVZ98,DZ93}. We have some progress in studying of a mixed problem where we come to a necessity of using of the declared matrix BA function. We believe that results of the paper will be useful for further development of the results obtained, for example, in \cite{GZM,Kis,K,M,MN} and for an investigation of rogue waves (about them see, e.g., \cite{ BET16, BG15, HC, MS16}) to the Maxwell--Bloch equations.

It is worth notice that it is very difficult to implement the algorithm \cite{BBEIM} (which uses a Riemann surface) for constructing the Baker--Akhiezer function associated with the Maxwell--Bloch system. The matter in fact of presence of a given broadening function $n(\la)$ is difficult to reconcile with the Riemann surface, which is the basic component of the method. To overcome this difficulty, it will be necessary to use Cauchy integrals with meromorphic/multi-valued kernels on the Riemann surface, which are very nontrivial for understanding to a wide range of specialists.

\section{Definition of the Baker--Akhiezer function and main results}\label{sec2}

In order to formulate our main results we start from the following definition of matrix Baker--Akhiezer function associated with the Maxwell--Bloch equations. First of all we fix the weight function $n(\la)$ ($\la\in\mathbb{R}$) which is smooth and satisfies~\eqref{n}. Let $\Sigma_j:=(E_j, E^*_j)$, $j=0, 1,2,\ldots,N$ be a set of vertical open intervals on the complex plane $\mathbb{C}$ which together with the real line $\mathbb{R}$ constitute an oriented contour $\Sigma=\mathbb{R} \cup\bigcup\limits_{j=0}^N\Sigma_j$. The orientation of $\mathbb{R}$ is chosen from left to right, and each $\Sigma_j$ is oriented from top to bottom (Fig.~\ref{fig1}). Boundary values of functions from the left and right of $\Sigma$ we denote by signs~$\pm$ respectively:
\begin{gather*}
\Psi_\pm(z)=\lim\limits_{z^\prime\to z\in\pm \text{side of }\Sigma}\Psi(z^\prime).
\end{gather*}
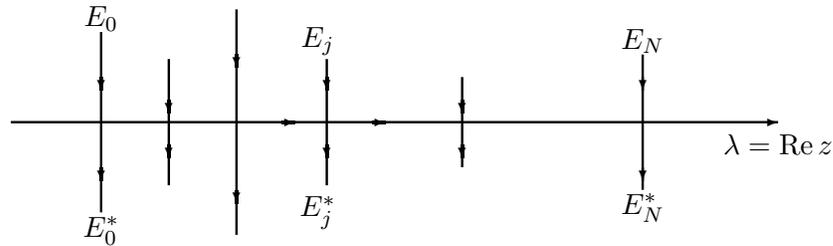
\begin{figure}[ht]
\vspace*{-23mm}
\begin{picture}(150,160)(-70,100)
\setlength{\unitlength}{0.60mm}
\linethickness{0,9pt}
\put(30.00,123.00){\makebox(0,0)[cc]{$E_0$}}\put(30.00,76.00){\makebox(0,0)[cc]{$E^*_0$}}
\put(30.00,120.00){\line(0,-1){40.00}}
\put(45.00,114.00){\line(0,-1){28.00}}
\put(60.00,125.00){\line(0,-1){50.00}}
\put(80.00,114.00){\line(0,-1){28.00}}
\put(110.00,110.00){\line(0,-1){20.00}}
\put(150.00,118.00){\makebox(0,0)[cc]{$E_N$}}\put(150.00,81.00){\makebox(0,0)[cc]{$E^*_N$}}
\put(150.00,115.00){\line(0,-1){30.00}}\put(150.00,110.0){\vector(0,-1){3.00}}\put(150.00, 90.0){\vector(0,-1){3.00}}
\put(180.00,95.00){\makebox(0,0)[cc]{$\la=\Re z$}}
\put(78.00,118.00){\makebox(0,0)[cc]{$E_j$}}
\put(78.00,80.00){\makebox(0,0)[cc]{$E^*_j$}}
\put(10.00,100.00){\vector(1,0){170.00}}
\linethickness{1,2pt}
\put(70.00,100.00){\vector(1,0){3.00}}\put(90.00,100.00){\vector(1,0){3.00}}
\put(30.00,110.0){\vector(0,-1){3.00}}\put(30.00,90.0){\vector(0,-1){3.00}}
\put(45.00,105.0){\vector(0,-1){3.00}}\put(45.00,95.0){\vector(0,-1){3.00}}
\put(60.00,115.0){\vector(0,-1){3.00}}\put(60.00,85.0){\vector(0,-1){3.00}}
\put(80.00,110.0){\vector(0,-1){3.00}}\put(80.00,95.0){\vector(0,-1){3.00}}
\put(110.00,105.0){\vector(0,-1){3.00}}\put(110.00,95.0){\vector(0,-1){3.00}}
\end{picture}
\vskip-1cm \caption{The oriented contour $\Sigma=\mathbb{R}\cup\bigcup\limits_{j=0}^N(E_j, E^*_j)$.}\label{fig1}
\end{figure}

\begin{Def}\label{def1} Let a contour $\Sigma$, a set of real constants $(\phi_0, \phi_1,\dots,\phi_N)$ and a weight function~$n(\la)$ be given. A $2\times2$ matrix $\Psi(t,x,z)$ is called the Baker--Akhiezer function associated with the Maxwell--Bloch equations if for any $x,t\in\mathbb{R}$:
\begin{itemize}\itemsep=0pt
\item $\Psi(t,x,z)$ is analytic in $z\in\mathbb{C}\setminus\overline\Sigma$, $\overline\Sigma:=\mathbb{R}\cup\bigcup\limits_{j=0}^N[E_j, E^*_j]$;
\item boundary values $\Psi_\pm(t,x,z)$ are continuous except for the endpoints $E_j$ and $E^*_j$, $j=0,1,\dots,N$ where $\Psi_\pm(t,x,z)$ have square integrable singularities;
\item boundary values $\Psi_\pm(t,x,z)$ are bounded at the points of self-intersection $\Re E_j$, $j=0,1,\dots,N$;
\item $\Psi(t,x,z)$ satisfies the jump conditions
\begin{gather*}
\Psi_-(t,x,z)=\Psi_+(t,x,z)J(x,z),\qquad z\in\Sigma,
\end{gather*}
where
\begin{gather}\label{RHPsi}
J(x,z) =\begin{pmatrix}
 \ee^{-\frac{\pi xn(\la)}{2}}& 0 \\
 0 & \ee^{\frac{\pi xn(\la)}{2}}
\end{pmatrix}, \qquad z=\la\in\mathbb{R}
\setminus\bigcup\limits_{j=0}^N\Re E_j,\\
 J(x,z) =\begin{pmatrix}
 0 & \ii e^{-\ii \phi_j}\\
 \ii e^{\ii \phi_j} & 0
 \end{pmatrix}, \qquad z\in\Sigma_j=(E_j, E^*_j), \qquad j=0,1,\dots,N;\label{ephij}
\end{gather}
\item $\Psi(t,x,z)$ satisfies the symmetry condition
\begin{gather*} \Psi(t,x,z)=\sigma_2\Psi^*(t,x,z^*)\sigma_2, \qquad
\sigma_2=\left(\begin{matrix}
 0 & -\ii \\
 \ii & 0
 \end{matrix}\right);\end{gather*}
\item $\Psi(t,x,z)=\big(I+O\big(z^{-1}\big)\big)e^{-\ii z(t-x)\sigma_3}$ as $z\to\infty$.
\end{itemize}
\end{Def}

These properties defines the matrix BA function uniquely and allow to construct $\Psi$ in an explicit form through theta functions and Cauchy integrals. To formulate main results let us define some necessary ingredients. Let
\begin{gather*}
w(z):=\sqrt{\prod_{j=0}^{N}(z-E_j)(z-E^*_j)}, \qquad
\varkappa(z):=\sqrt[4]{\prod\limits_{j=0}^{N} \frac{z-E^*_j}{z- E_j}},
\qquad z\in\mathbb{C}\setminus\bigcup\limits_{j=0}^N [E_j, E^*_j]
\end{gather*}
be roots whose branches are fixed by cuts along $[E_j, E^*_j]$, $j=0,\dots, N$ and conditions $w(z)\simeq z^{N+1}$, $\varkappa(z)\simeq 1 $ as $z\to\infty$. Define scalar functions $f(z)$ and $g(z)$ through Cauchy integrals
\begin{gather}\label{f-gen}
f(z) = \frac{w(z)}{2\pi \ii} \sum_{j=1}^{N}\int_{\Sigma_j}\frac{C^f_j}{w_+(\xi)(\xi-z)}\dd\xi,\\
\label{g-gen-n}
g(z) = \frac{w(z)}{2\pi \ii}\sum_{j=1}^{N}\int_{\Sigma_j}\frac{C^g_j}{w_+(\xi)(\xi-z)}\dd\xi+
\frac{w(z)}{4}\int_{\mathbb{R}}\frac{n(\la)}{w(\la)(\la-z)}\dd\la,
\end{gather}
where $ C^f_j$, $ C^g_j$ are uniquely defined by linear algebraic equations
\begin{gather}
\sum_{j=1}^N C^f_j\int_{\Sigma_j}\frac{\xi^k \dd\xi}{w_+(\xi)} = 0, 		 \qquad k=0,\dots,N-2, \nonumber\\
\sum_{j=1}^N C^f_j\int_{\Sigma_j}\frac{\xi^{N-1} \dd\xi}{w_+(\xi)} = -2\pi\ii, \label{Cf}\\
\sum_{j=1}^N C^g_j\int_{\Sigma_j}\frac{\xi^k \dd\xi}{w_+(\xi)} = -\frac{\ii\pi}{2}\int_{\mathbb{R}}\frac{\la^k n(\la)}{w(\la)}\dd\la, 		 \qquad k=0,\dots,N-2,\nonumber \\
\sum_{j=1}^N C^g_j\int_{\Sigma_j}\frac{\xi^{N-1} \dd\xi}{w_+(\xi)}=
2\pi\ii -\frac{\ii\pi}{2}\int_{\mathbb{R}}\frac{\la^{N-1} n(\la)}{w(\la)}\dd\la. \label{Cg}
\end{gather}
A unique solvability of \eqref{Cf} and \eqref{Cg} is well-known. A detailed proof can be found in \cite[Problem~9.4.2, pp.~234--235]{TDT2013} or~\cite{TO16}.

\begin{thm}\label{T1}
Let a contour $\Sigma$, a set of real constants $(\phi_0, \phi_1,\dots,\phi_N)$ and a weight $($smooth$)$ function $n(\la)$ be given. Let all requirements of Definition~{\rm \ref{def1}} are fulfilled. Then $\Psi$ is unique and takes the form
\begin{gather*}
\Psi(t,x,z) = \ee^{(\ii t f_0 +\ii x g_0 )\sigma_3} M(t,x,z) \ee^{-(\ii tf(z) +\ii xg(z) )\sigma_3},
\end{gather*}
where constants $f_0$ and $g_0$ are equal to
\begin{gather}\label{f0}
f_0=-\sum\limits_{j=0}^N\Re E_j- \frac{1}{2\pi \ii} \sum_{j=1}^{N}\int_{\Sigma_j}\frac{C^f_j \xi^N}{w_+(\xi)}\dd\xi,
\\
\label{g0}
g_0=\sum\limits_{j=1}^N\Re E_j- \frac{1}{2\pi \ii} \sum_{j=1}^{N}\int_{\Sigma_j}\frac{C^g_j \xi^N}{w_+(\xi)}\dd \xi -
\frac{1}{4}\int_{\mathbb{R}}\frac{\la^N n(\la)}{w(\la)}\dd\la,
\end{gather}
functions $f(z)$ and $g(z)$ are given by \eqref{f-gen}--\eqref{Cg}, and $M(t,x,z)$ is a solution of the following RH problem:
\begin{itemize}\itemsep=0pt
\item $M(t,x,z)$ is analytic in $z\in\mathbb{C}\setminus\bigcup\limits_{j=0}^N[E_j, E^*_j]$;
\item boundary values $M_\pm(t,x,z)$ are continuous, except for the endpoints $E_j$ and $E^*_j$ where $M_\pm$ have square integrable singularities;
\item $M(t,x,z)$ satisfies the jump conditions
\begin{gather}\label{RHM}
M_-(t,x,z)=M_+(t,x,z)J_M(t,x,z),\qquad z\in\Sigma_j=(E_j, E^*_j),\quad j=0,1,\dots,N,\!\!\!
\\
\label{JM}
J_M(t,x,z)=\begin{pmatrix}
 0 & \ii e^{-\ii (t C^f_j+x C^g_j+\phi_j)}\\
 \ii e^{\ii (t C^f_j+x C^g_j+\phi_j)} & 0
 \end{pmatrix}, \qquad z\in\Sigma_j=(E_j, E^*_j);
\end{gather}
\item $M(t,x,z)$ satisfies the symmetry condition $M(t,x,z)=\sigma_2M^*(t,x,z^*)\sigma_2$;
\item $M(t,x,z)=I+O\big(z^{-1}\big) $ as $z\to\infty$.
\end{itemize}
\end{thm}
The next theorem presents an explicit formula for $M(t,x,z)$.
\begin{thm}\label{MinTheta} Under conditions of the Theorem~{\rm \ref{T1}} entries of matrix $M(t,x,z)$ are
\begin{gather*}
M_{11}(t,x,z)=\frac{\varkappa(z)+\varkappa^{-1}(z)}{2} \frac{\Theta(\B A(\infty)+\B A(\mathcal{D})+\B K)}
{\Theta(\B A(z)+\B A(\mathcal{D})+\B K)} \frac{\Theta(\B A(z)+\B A(\mathcal{D})+\B K+ \B C(t,x))}
{\Theta(\B A(\infty)+\B A(\mathcal{D})+\B K+\B C(t,x))},\\
M_{12}(t,x,z)=\frac{\varkappa(z)-\varkappa^{-1}(z)}{2}\ee^{-\ii\phi_0} \frac{\Theta(\B A(\infty)+\B A(\mathcal{D})+\B K)
\Theta(\B A(z)-\B A(\mathcal{D})-\B K- \B C(t,x))}
{\Theta(\B A(z)-\B A(\mathcal{D})-\B K)\Theta(\B A(\infty)+\B A(\mathcal{D})+\B K+\B C(t,x))},\\
M_{21}(t,x,z)=\frac{\varkappa(z)-\varkappa^{-1}(z)}{2} \ee^{\ii\phi_0} \frac{\Theta(\B A(\infty)+\B A(\mathcal{D})+\B K)}
{\Theta(\B A(z)-\B A(\mathcal{D})-\B K)} \frac{\Theta(\B A(z)-\B A(\mathcal{D})-\B K+ \B C(t,x))}
{\Theta(\B A(\infty)+\B A(\mathcal{D})+\B K-\B C(t,x))}, \!\\
M_{22}(t,x,z)=\frac{\varkappa(z)+\varkappa^{-1}(z)}{2} \frac{\Theta(\B A(\infty)+\B A(\mathcal{D})+\B K)}
{\Theta(\B A(z)+\B A(\mathcal{D})+\B K)} \frac{\Theta(\B A(z)+\B A(\mathcal{D})+\B K-\B C(t,x))}
{\Theta(\B A(\infty)+\B A(\mathcal{D})+\B K-\B C(t,x))},
\end{gather*}
where $\Theta$ is theta-function \eqref{theta} defined by the Fourier series
\begin{gather*}
\Theta(\B u)=\sum\limits_{\B{l}\in\mathbb{Z}^N} \exp{\{ \pi\ii(B\B{l},\B{l})+2\pi\ii (\B{l},\B{u}) \}}, \qquad (\B{l},\B{u})= l_1u_1+\dots+l_Nu_N,
\end{gather*}
and $\B A(z)$, $\B A(\mathcal{D})$ are Abel mapping \eqref{AP}, \eqref{AD}, $\B K$ is a vector of Riemann constants \eqref{K}. The dependence of $M(t,x,z)$ in $t$ and $x$ is determined by vector-function with components
\begin{gather*}
{C}_j(t,x):=- \frac{t{C^f_j} +x{ C^g_j}+{\mathbb\phi_j}}{2\pi},\qquad j= 1, 2, \dots, N.
\end{gather*}
\end{thm}

\begin{thm}\label{PsiAKNS} Let $\Psi$ is defined by Theorems~{\rm \ref{T1}} and~{\rm \ref{MinTheta}}. Then for any $z\in\mathbb{C}\setminus\Sigma$ matrix $\Psi(t,x,z)$ is smooth in $t$ and $x$ and satisfies AKNS equations
\begin{gather}\label{PsitPsix}
\Psi_t = -(\ii z\sigma_3+H(t,x))\Psi, \qquad
\Psi_x =(\ii z\sigma_3+H(t,x)-\ii G(t,x,z))\Psi,
\end{gather}
where $H(t,x)$ is given by
\begin{gather}\label{H}
H(t,x)=-\ii\ee^{\ii(tf_0+xg_0)\sigma_3}[\sigma_3, m(t,x)]\ee^{-\ii(tf_0+xg_0)\sigma_3},\\
 m(t,x)= \lim\limits_{z\to\infty} z (M(t,x,z) - I),\nonumber
\end{gather}
and
\begin{gather*}
G(t,x,z)= \frac{1}{4}\int_{-\infty}^\infty \frac{F(t,x,s)n(s)}{s-z}\dd s, \qquad z\notin\mathbb{R}.
\end{gather*}
Matrix $F(t,x,\lambda)$ is Hermitian, has unit determinant and presented by formula
\begin{gather}\label{FviaM}
F(t,x,\lambda)=\ee^{\ii(tf_0+xg_0)\sigma_3}M(t,x,\la)\sigma_3M^{-1}(t,x,\la)\ee^{-\ii(tf_0+xg_0)\sigma_3},\\
 \la\neq\Re E_j, \qquad j= 0, 1, 2, \dots, N.\nonumber
\end{gather}
\end{thm}

\begin{thm}\label{fgsol}
The associated with $\Psi(t,x,z)$ finite-gap solution to the Maxwell--Bloch equations \eqref{MB1}--\eqref{MB3} is given by
\begin{gather}
{\mathcal E}(t,x)=E_{\Theta} \frac{\Theta(-\B A(\infty)+\B A(\mathcal{D})+\B K+\B C(t,x))}{\Theta(\B A(\infty)+\B A(\mathcal{D})+\B K+\B C(t,x))}
\ee^{2\ii(tf_0+xg_0)-\ii\phi_0},\label{solMBE}
\end{gather}
where
\begin{gather*}
E_{\Theta}:=2\frac{\Theta(\B A(\infty)+\B A(\mathcal{D})+\B K)}{\Theta(-\B A(\infty)+\B A(\mathcal{D})+\B K)} \sum^{N}_{j=0}\Im E_j,
\end{gather*}
$f_0$ and $g_0$ are defined by \eqref{f0} and \eqref{g0}. The dependence of the solution in $t$ and $x$ is determined by the $N$ dimensional $($linear in $t$ and $x)$ vector-function
\begin{gather*}
{\B C}(t,x):=- \frac{t{\B C^f} +x{\B C^g}+{\mathbb\phi}}{2\pi}.
\end{gather*}
The density matrix $F(t,x,\la)$ equals to
\begin{gather}\label{solF}
\begin{pmatrix}{\mathcal N}(t,x,\la)&{\mathcal\rho}(t,x,\la)\\
{\mathcal\rho}^*(t,x,\la)&-{\mathcal N}(t,x,\la)\end{pmatrix}
= \ee^{\ii(tf_0+xg_0)\sigma_3}M(t,x,\la)\sigma_3 M^{-1}(t,x,\la)\ee^{-\ii(tf_0+xg_0)\sigma_3}.
\end{gather}
Moreover, the finite-gap solution $\mathcal{E}(t,x)$, $\mathcal N(t,x,\la))$, $\rho(t,x,\la)$ to the Maxwell--Bloch equations \eqref{MB1}--\eqref{MB3} are smooth for $t,x,\la \in\mathbb{R}$, except for the $\la_j:=\Re E_j$, $j=0,1,2,\ldots,N$.
\end{thm}

The paper is organized as follows. In Section~\ref{sec3}, we prove the Theorem~\ref{T1}. In Section~\ref{sec4}, we give a construction of the phases~$f$ and~$g$ by Cauchy integrals, and in Section~\ref{sec5}, we propose another representations for them using hyperelliptic integrals. In Section~\ref{sec6}, explicit construction of $M(t,x,z)$ is presented (the proof of Theorem~\ref{MinTheta}). In Section~\ref{sec7}, we deduce AKNS equations for $\Psi(t,x,z)$ (the proof of Theorem~\ref{PsiAKNS}). Section~\ref{sec8} describes finite-gap solutions to the MB equations (the proof of the Theorem~\ref{fgsol}). Section~\ref{sec9} contains final remarks.

\section[Proof of the Theorem \ref{T1} and RH problem for $M=M(t,x,z)$]{Proof of the Theorem \ref{T1} and RH problem for $\boldsymbol{M=M(t,x,z)}$}\label{sec3}

\begin{proof}[Uniqueness] The matrix $\Psi(t,x,z)$ has unit determinant. Indeed, since $\Psi$ is a matrix of the second order then, due to definition of $\Psi$, $\det\Psi$ is analytic in $z\in\mathbb{C}\setminus\Sigma$, continuous up to the contour $\Sigma$, except for the endpoints $E_j$, $E_j^*$ where it has weak singularities, and bounded at all self-intersection points $\Re E_j$. In view of~\eqref{RHPsi}, $\det J(x,z)\equiv1$, hence
 \begin{gather*}
\det\Psi_-(t,x,z)=\det\Psi_+(t,x,z), \qquad z\in\Sigma,
\end{gather*}
i.e., $\det\Psi$ has no jump at the contour $\Sigma$. Therefore $\det\Psi$ is analytic everywhere, except for a set of self-intersection points and endpoints of $\Sigma$ where it has removable singularities. At infinity $\det\Psi(t,x,z)=1+\ord\big(z^{-1}\big)$, hence $\det\Psi(t,x,z)\equiv 1$ by Liouville theorem. In particular, $\Psi(t,x,z)$ is invertible for any $z$ outside the exceptional set. Suppose that $\tilde \Psi(t,x,z)$ is another solution of the RH problem. Then $\Phi(z):=\tilde \Psi(t,x,z)\Psi^{-1}(t,x,z)$
satisfies
\begin{gather*}
\Phi_-(z)=\tilde \Psi_-(t,x,z)\Psi_-^{-1}(t,x,z)=\tilde \Psi_+(t,x,z)J(x,z)J^{-1}(x,z)\Psi_+^{-1}(t,x,z)=\Phi_+(z),
\end{gather*}
and \looseness=1 it is continuous across $\Sigma$ with exception of end points $E_j$, $E^*_j$ and points of self-intersection $\Re E_j$. These points are removable singularities. Hence $\Phi(t,x,z)$ has an analytic continuation for $z\in\mathbb{C}$ and it tends to identity matrix as $z\to\infty$. By Liovilles's theorem $\Phi(t,x,z)=\tilde\Psi(t,x,z)\Psi^{-1}(t,x,z)\equiv I$ and therefore $\tilde \Psi(t,x,z)\equiv\Psi(t,x,z)$, i.e., the matrix $\Psi(t,x,z)$ is unique.
\end{proof}

\begin{proof}[Existence]
To prove the existence of the Baker--Akhiezer function we use an explicit construction of $\Psi$ using different RH problems. To transform the initial RH problem to a form allowing an explicit solution, let us seek $\Psi(t,x,z)$ in the form
\begin{gather}\label{PsiM}
\Psi(t,x, z) = \ee^{\ii (tf_0 + g_0 x)\sigma_3} M(t,x,z) \ee^{-\ii (tf(z) +g(z)x )\sigma_3},
\end{gather}
where constants $f_0$ and $g_0$, scalar functions $f(z)$ and $g(z)$ and matrix $M(t,x,z)$ are to be determined. The symmetry of $\Psi$ \eqref{symPsi} produces symmetries of $f(z)$ and $g(z)$, i.e., they have to satisfy the conditions: $f^*(z^*)= f(z)$ and $g^*(z^*)= g(z)$, particularly $f^*_0= f_0$ and $g^*_0= g_0$.

Due to the definition of $\Psi$ we obtain the RH problem \eqref{RHM}, \eqref{JM}. Indeed, all above statements will be true if $f(z)$ and $g(z)$ possess properties:
\begin{itemize}\itemsep=0pt	
\item $f(z) $ is analytic in $z\in\mathbb{C}\setminus\bigcup\limits_{j=0}^N[E_j, E^*_j]$;
\item $f(z)=f^*(z^*)$ and
\begin{gather}\label{f}	
f(z)=z+f_0+O(1/z), \qquad\text{as}\quad z\to\infty;
\end{gather}
\item $f_+(z)+f_-(z)=C^f_j$, $z\in\Sigma_j$, $j=0,1,\dots,N$,
\end{itemize}
where $f_0$ and $C^f_j$ are some real (as a result of the symmetry of $\Psi$) constants;
\begin{itemize}\itemsep=0pt	
\item $g(z) $ is analytic in $z\in\mathbb{C}\setminus(\mathbb{R}\cup\bigcup\limits_{j=0}^N[E_j, E^*_j])$;
\item $g(z)=g^*(z^*)$ and
\begin{gather}\label{g}
 	g(z)=-z+g_0+O(1/z), \qquad \text{as}\quad z\to\infty;
\end{gather}
\item $g_+(z)+g_-(z)=C^g_j$, $z\in\Sigma_j$, $j=0,1,\dots,N$;
\item $g_+(\la)-g_-(\la)= \frac{\pi\ii}{2}n(\la)$, $\la\in\mathbb{R}\setminus\bigcup\limits_{j=0}^N\Re E_j$,
\end{itemize}
where $g_0$ and $C^g_j$ are some real constants. All constants $f_0$, $g_0$, $C^f_j$, $C^g_j$, $j=0,1,2,\ldots, N$, are determined in the next section where we prove formulas \eqref{f-gen}--\eqref{g0}.

Asymptotics \eqref{f}, \eqref{g} give that $M(t,x,z)=I+O\big(z^{-1}\big)$ as $z\to\infty$. The jumps of func\-tions~$f(z)$ and~$g(z)$ provide the form of matrix~\eqref{JM}. Indeed, for $z\in\Sigma_j$,
\begin{gather*}
J_M(t,x,z)=\ee^{-\ii(tf_+(z)+g_+(z) x)\sigma_3}J(x,z)\ee^{\ii(tf_-(z)+g_-(z) x)\sigma_3}\\
 \hphantom{J_M(t,x,z)}{} =\begin{pmatrix}
 0& \ii e^{-\ii t (f_+(z) +f_-(z)) -\ii x(g_+(z)+g_-(z))-\ii\phi_j)}\\
 \ii e^{\ii t (f_+(z) +f_-(z)) +\ii x(g_+(z)+g_-(z))+\ii\phi_j)} & 0
 \end{pmatrix}
\end{gather*}
that gives \eqref{JM}. We stress that such a choice of the jump matrices on intervals $\Sigma_j$ (independent on~$z$) provides solvability of the RH problem for $M(t,x,z)$ in an explicit form in theta functions.

The jump of the function $g(z)$ on the real axis ($z=\la$) makes the matrix $M(t,x,z)$ to be continuous on $\mathbb{R}\setminus\bigcup\limits_{j=0}^N\Re E_j$:
\begin{gather*}
J_M(t,x,z)=\ee^{-\ii(tf_+(z)+g_+(z) x)\sigma_3}J(x,z)\ee^{\ii(tf_-(z)+g_-(z) x)\sigma_3}\\
\hphantom{J_M(t,x,z)}{} =\ee^{-\ii t(f_+(z)-f_-(z) )\sigma_3}\ee^{-\frac{\pi xn(\la)\sigma_3}{2}}\ee^{-\ii x(g_+(z)-g_-(z) )\sigma_3}\\
\hphantom{J_M(t,x,z)}{} = \ee^{-\frac{\pi xn(\la)\sigma_3}{2}}\ee^{\frac{\pi xn(\la)\sigma_3}{2}}=I,\qquad
 z=\la\in\mathbb{R}\setminus\bigcup\limits_{j=0}^N\Re E_j
\end{gather*}
and thus $M$ is analytic in $z\in\mathbb{C}\setminus\bigcup\limits_{j=0}^N[E_j, E^*_j]$. The symmetries of $M=M(z)$ follow from the symmetry of jump contour $\Sigma$ with respect to the real axis and symmetric properties of scalar functions $f(z)$ and $g(z)$ and matrix $\Psi=\Psi(z)$. Finally, it is important to emphasize a~normalization condition
\begin{gather*}
\det\Psi(t,x,z)=\det M(t,x,z)\equiv 1,
\end{gather*}
which follows from the definition of $\Psi$.
\end{proof}

The Riemann--Hilbert problems like \eqref{JM} have already been encountered in different form in the so-called model problems (see, for example, publications \cite{BT16,BK07a,BIK09,BK07,BKS11,BV07,BM13,BM14,D,DIZ93,DIZ97, DKMVZ99-1,DKMVZ99-2,DVZ98,DZ93,EGKT13,KMM03,KSZ15,KT12,KM10,KM12, MK06,MK10,TVZ04,TVZ06,TO16}). All these papers were devoted to studying an asymptotic behavior of different problems arising in the soliton theory, in the theory of random matrix models, and also in the theory of integrable statistical mechanics. These model problems have auxiliary in nature, and for our constructions it is impossible to use the results of those articles directly. Therefore for the completeness of exposition we give in the next sections an explicit construction of scalar functions $f(z)$, $g(z)$ and matrix $M(t,x,z)$ by using ideas of just cited articles and also of the paper~\cite{KS}.

\section[Construction of the phases $f$ and $g$ by Cauchy integrals]{Construction of the phases $\boldsymbol{f}$ and $\boldsymbol{g}$ by Cauchy integrals}\label{sec4}

In this section we give the construction of the phase functions $f$ and $g$. We start from the case involving only one arc. In this case, the jump conditions for~$f$ and~$g$ are jumps of type \eqref{f} and \eqref{g}{\samepage
\begin{gather} \label{f-g-jump-gen0}
f_+(z)+f_-(z)=C^f_0, \qquad g_+(z)+g_-(z)=C^g_0
\end{gather}
across a single arc $\Sigma_0$.}

Define
\begin{gather*}
w(z):=\sqrt{(z-E_0)(z-E^*_0)}
\end{gather*}
such that $w(z)$ is analytic outside the arc and $w(z)\simeq z$ as $z\to\infty$, and introduce
\begin{gather} \label{tilde}
\tilde f:=\frac{f}{w}, \qquad \tilde g:=\frac{g}{w}.
\end{gather}
Then the jump conditions (\ref{f-g-jump-gen0}) reduce to
\begin{gather*}
\tilde f_+(z)-\tilde f_-(z)= \frac{C^f_0}{w_+}, \qquad \tilde g_+(z)-\tilde g_-(z)=\frac{C^g_0}{w_+}, \qquad z\in \Sigma_0,\\
\tilde g_+(\la)-\tilde g_-(\la)= \frac{\ii \pi n(\la)}{2w(\la)}, \qquad \la\in\mathbb{R}\setminus\{\Re E_0\}.
\end{gather*}
Due to the asymptotic conditions (\ref{f}), $\tilde f =1 +O(1/z)$ as $z\to\infty$, and thus $\tilde f$ is (uniquely) determined by $C^f_0$ through Cauchy integral
\begin{gather*}
\tilde f(z) = 1+\frac{1}{2\pi \ii}\int_{\Sigma_0}\frac{C^f_0}{w_+(\xi)(\xi-z)}\dd\xi=1+\frac{C^f_0}{2w(z)}.
\end{gather*}
Consequently,
\begin{gather*}
f(z) = w(z)\left(1+\frac{1}{2\pi \ii}\int_{\Sigma_0}\frac{C^f_0}{w_+(\xi)(\xi-z)}\dd\xi \right)=w(z)+\frac{C^f_0}{2}.
\end{gather*}
Particularly, $f_0$ is determined by
\begin{gather*}
f_0 =- \frac{1}{2\pi \ii}\int_{\Sigma_0}\frac{C^f_0}{w_+(\xi)}\dd\xi-\frac{1}{2}(E_0+E^*_0)=-\Re E_0+\frac{C^f_0}{2}.
\end{gather*}
Taking into account (\ref{PsiM}) it can be put $C^f_0=0$ without loss of generality and hence $f(z)=w(z)=\sqrt{(z-E_0)(z- E^*_0)}$
and $f_0=-\Re E_0$.

Now consider the function $g(z)$. In this case we have $\tilde g = g w^{-1} =-1 +O(1/z)$ as $z\to\infty$, and thus
\begin{gather*}
\tilde g(z) = -1+ \frac{1}{2\pi \ii}\int_{\Sigma_0}\frac{C^g_0}{w_+(\xi)(\xi-z)}\dd\xi+ \frac{1}{4}\int_{\mathbb{R}}\frac{n(\la)}{w(\la)(\la-z)}\dd\la\\
\hphantom{\tilde g(z)}{} =-1+\frac{1}{4}\int_{\mathbb{R}}\frac{n(\la)}{w(\la)(\la-z)}\dd\la +\frac{C^g_0}{2}.
\end{gather*}
Consequently, by the same reason as above with $C^g_0=0$
\begin{gather*}
g(z) = -w(z)\left(1-\frac{1}{4}\int_{\mathbb{R}}\frac{n(\la)}{w(\la)(\la-z)}\dd\la\right),
\end{gather*}
and, particularly,
\begin{gather*}
g_0 = \Re E_0 -\frac{1}{4}\int_{\mathbb{R}}\frac{n(\la)}{w(\la)}\dd\la.
\end{gather*}

Now consider the general case, where the contour consists of $N+1$, $N\ge 1$, arcs $\Sigma_j$, $j=0,\dots, N$. Define
\begin{gather*}
w(z):=\sqrt{\prod_{j=0}^{N}(z-E_j)(z-E^*_j)}
\end{gather*}
such that $w(z)$ is analytic outside the arcs $\Sigma_j$ and $w(z)\simeq z^{N+1}$ as $z\to\infty$, and introduce~$\tilde f$ and~$\tilde g$ as in~(\ref{tilde}). The jump conditions reduce to
\begin{gather}\label{tilde-f-g}
\tilde f_+(z)-\tilde f_-(z)=\frac{C^f_j}{w_+}, \qquad \tilde g_+(z)-\tilde g_-(z)=\frac{C^g_j}{w_+}, \qquad z\in\Sigma_j,\\
\label{tilde-g}
\tilde g_+(z)-\tilde g_-(z)= \frac{\ii \pi n(\la)}{2w(\la)}, \qquad \la\in\mathbb{R}\setminus\bigcup\limits_{j=0}^N\{\Re E_j\}.
\end{gather}

For $N\ge 1$ we have
\begin{gather*}
f(z) = \frac{w(z)}{2\pi \ii} \sum_{j=0}^{N}\int_{\Sigma_j}\frac{C^f_j}{w_+(\xi)(\xi-z)}\dd\xi\\
\hphantom{f(z)}{} = \frac{w(z)}{2\pi \ii} \sum_{j=0}^{N}\left(\int_{\Sigma_j}\frac{C^f_0}{w_+(\xi)(\xi-z)}\dd\xi+\int_{\Sigma_j}\frac{C^f_j-C^f_0}{w_+(\xi)(\xi-z)}\dd\xi\right).
\end{gather*}
Since $w(z)$ is analytic in $z\in \mathbb{C}\setminus\bigcup\limits_{j=0}^{N}\overline{\Sigma_j}$ Cauchy theorem gives
\begin{gather*}
\frac{1}{2w(z)}=\frac{1}{2\pi \ii} \sum_{j=0}^{N}\int_{\Sigma_j}\frac{1}{w_+(\xi)(\xi-z)}\dd\xi.
\end{gather*}
Hence
\begin{gather*}
f(z) = \frac{C^f_0}{2}+\frac{w(z)}{2\pi \ii} \sum_{j=0}^{N}\int_{\Sigma_j}\frac{C^f_j-C^f_0}{w_+(\xi)(\xi-z)}\dd\xi.
\end{gather*}
Again, it is convenient to put $C^f_0=0$. Then, in view of $\tilde f = O(1/z)$ as $z\to\infty$, $C^f_j$ have to satisfy the system of linear equations
\begin{gather*}
\sum_{j=1}^{N}\int_{\Sigma_j}\frac{\xi^m C^f_j}{w_+(\xi)}\dd\xi=-2\pi\ii\delta_{m,N-1}
\end{gather*}
for $m=0,1,2,\ldots,N-1$. Thus we have $N$ equations \eqref{Cf} for $N$ unknown constants $C^f_j$. It is well known (see, for example, \cite{BBEIM,D1,TO16,Zver}) that this system of linear algebraic equations has a unique solution $\{C^f_j\}_{j=1}^N$. Then $f(z)$ and $f_0$ takes the form~\eqref{f-gen} and~\eqref{f0}.

Now consider the function $g(z)$. In view of \eqref{tilde-f-g}, \eqref{tilde-g}, and take into account that for $N\ge 1$ $\tilde g = O(1/z)$ as $z\to\infty$ we have
\begin{gather*}
\tilde g(z) = \frac{1}{2\pi \ii} \sum_{j=0}^{N}\int_{\Sigma_j}\frac{C^g_j}{w_+(\xi)(\xi-z)}\dd\xi +
\frac{1}{4}\int_{\mathbb{R}}\frac{n(\la)}{w(\la)(\la-z)}\dd\la,
\end{gather*}
where, by the same reasons as above we put $C^g_0=0$. Consequently, $g(z)$ takes the form \eqref{g-gen-n}. Now the requirement that $g(z)$ given by (\ref{g-gen-n}) satisfies the asymptotic condition (\ref{g}) leads to a~system of~$N$ linear algebraic equation for $C^g_j$, $j=1,\dots,N$. Indeed, if we use the asymptotic for large $z$ expansion
\begin{gather*}
\frac{1}{2\pi \ii} \sum_{j=1}^{N}\int_{\Sigma_j}\frac{C^g_j}{w_+(\xi)(\xi-z)}\dd\xi +
\frac{1}{4}\int_{\mathbb{R}}\frac{n(\la)}{w(\la)(\la-z)}\dd\la =\sum\limits_{l=0}^\infty \frac{I_l}{z^{l+1}},
\end{gather*}
then, due to (\ref{g}), it is evident that $I_0=I_1=\dots=I_{N-2}=0$. Hence
\begin{gather*}
g(z)=-\big(z^{N+1}+w_Nz^N+ \cdots \big) \left(\frac{I_{N-1}}{z^N}+\frac{I_N}{z^{N+1}}+\cdots\right)\\
\hphantom{g(z)}{} =-\big(zI_{N-1}+I_{N-1}w_N+I_N+ \ord\big(z^{-1}\big) \big)= -z+g_0 + \ord\big(z^{-1}\big)
\end{gather*}
and thus $I_{N-1}=1$, and $g_0=-w_N-I_N$ where $w_N=-\sum\limits_{j=0}^N\Re E_j$. This gives the system of linear algebraic equations~\eqref{Cg}. Similarly to (\ref{Cf}), (\ref{Cg}) has a unique solution. A detailed proof can be found in~\cite[Problem~9.4.2, pp.~234--235]{TDT2013} or in~\cite{TO16}. The parameter~$g_0$ is given by~\eqref{g0}.

\section[Representation of $f(z)$ and $g(z)$ through hyperelliptic integrals]{Representation of $\boldsymbol{f(z)}$ and $\boldsymbol{g(z)}$\\ through hyperelliptic integrals}\label{sec5}

Here we give another representation of $f(z)$ and $g(z)$, using hyperelliptic integrals. We seek $f(z)$ in the form
\begin{gather*}
f(z)=\int_{E^*_{0}}^z\varphi(\la) \dd\la\qquad \text{with}\quad \varphi(\la)= \frac {\hat f(\la)}{w(\la)},
\end{gather*}
where $w^2(\la)= \prod\limits_{j=0}^{N}(\la-E_j)(\la- E^*_j) \equiv \la^{2(N+1)}+P_{2N+1}\la^{2N+1}+P_{2N}\la^{2N}+\dots +P_0=P(\la)$
and $\hat f(\la)=\la^{N+1}+\hat f_N\la^N+\hat f_{N-1}\la^{N-1}+\dots +\hat f_0$. Asymptotics (\ref{f}) of the function $f(z)$ defines $\hat f_N=\frac{P_{2N+1}}{2}$. In order to define $\hat f_0, \hat f_1, \dots, \hat f_{N-1}$, we normalize $f(z)$ by the conditions
\begin{gather*}
\int_{E_{j}}^{E^*_j}\dd f=0, \qquad j=1,\dots,N.
\end{gather*}
In other words, for $z\in\mathbb{C}\setminus\bigcup\limits_{j=0}^N [E_j, E_j^*] $ the function $f(z)$ can be considered as a hyperelliptic integral of the second kind with simple pole at infinity. Integral $f(z)$ is uniquely fixed by the condition of zero $a$-periods~\cite{BBEIM}. They are $A^f_j=2\int_{E_{j}}^{E^*_j}\dd f=0$. Indeed, since $\varphi_+(\la)+\varphi_-(\la)=0$ for $z\in [E_j, E_j^*]$ and $\varphi_+(\la)+\varphi_-(\la)=2\varphi(\la)$ for $z\in [\Re E_j, \Re E_{j+1}]$, it is easy to check that $C_0^f =\int_{ E^*_{0}}^z(\varphi_+(\la)+\varphi_-(\la)) \dd\la=0$ whereas $C^f_j$ for $j>0$ are determined as the (nonzero) $b$-periods of~$f(z)$
\begin{gather}
C^f_j=f_+(z)+f_-(z) =\int_{E^*_{0}}^{ z}(\varphi_+(\la)+\varphi_-(\la))\dd\la\nonumber \\
\hphantom{C^f_j=f_+(z)+f_-(z)}{} =2\sum\limits_{l=1}^j\int_{E^*_{l-1}}^{E_l}\varphi(\la)\dd\la =:B^f_j\neq0,\qquad j=1,\dots,N,\label{C=B}
\end{gather}
where $B^f_j=\int_{\B{b}_j} \dd f(z)$. The last equality becomes obvious if we use the definition of $\B a$- and $\B b$-cycles of the hyperelliptic surface given by the function $w(z)$ (see the next section and Fig.~\ref{fig2}). On the other hand, for $z\in\mathbb{R}\setminus\bigcup\limits_{j=0}^N\{\Re E_j\}$
\begin{gather*}
f_+(z)-f_-(z)=\int_{E^*_{0}}^{ z}(\varphi_+(\la)-\varphi_-(\la))\dd\la\\
\hphantom{f_+(z)-f_-(z)}{}=2\sum\limits_{l=1}^j\int_{E_{l}}^{ E^*_{l}}\varphi_+(\la)\dd\la =\sum\limits_{l=1}^j A^f_l=0, \qquad j=1,\dots,N.
\end{gather*}

The function $g(z)$ cannot be written as a hyperelliptic integral, but it is determined as a sum of the hyperelliptic integral $-f(z)$ and Cauchy integrals
\begin{gather}\label{g-hyper}
g(z) = -f(z)+\frac{w(z)}{2\pi \ii}\sum_{j=0}^{N}\int_{\Sigma_j}\frac{C^h_j}{w_+(\xi)(\xi-z)}\dd\xi+
\frac{w(z)}{4}\int_{\mathbb{R}}\frac{n(\la)}{w(\la)(\la-z)}\dd\la,
\end{gather}
where constants $\{C^h_j\}_{j=1}^N$ have to be determined. To prove this formula let us put
\begin{gather*}
h(z):= \frac{f(z)+g(z)}{w(z)}.
\end{gather*}
Equations \eqref{f} and \eqref{g} provide the following properties of function $h(z)$:
\begin{itemize}\itemsep=0pt	
\item $h(z) $ is analytic in $z\in\mathbb{C}\setminus\big(\mathbb{R}\cup\bigcup\limits_{j=0}^N[E_j, E^*_j]\big)$;
\item $	h(z)=O(1/z)$, as $z\to\infty$;
\item $h_+(z)-h_-(z)= \frac{C^h_j}{w(z)}$, $z\in\Sigma_j$, $j=0,1,\ldots,N$;
\item $h_+(\la)-h_-(\la)= \frac{\pi\ii}{2w(\la)}n(\la)$, $\la\in\mathbb{R}\setminus\bigcup\limits_{j=0}^N\Re E_j$,
\end{itemize}
where $C^h_j=C^g_j+ C^f_j$ are to be determined. Due to \eqref{C=B} $C^f_j$ are already known: $C^f_j=B^f_j$, $j=1,\dots,N$. Then $h(z)$ can be written as a sum of Cauchy integrals
\begin{gather*}
h(z) = \frac{1}{2\pi \ii}\sum_{j=0}^{N}\int_{\Sigma_j}\frac{C^h_j}{w_+(\xi)(\xi-z)}\dd\xi+
\frac{1}{4}\int_{\mathbb{R}}\frac{n(\la)}{w(\la)(\la-z)}\dd\la,
\end{gather*}
and hence \eqref{g-hyper} follows. The asymptotic condition (\ref{g}) leads to a system of $N$ linear equation for $C^h_j$, $j=1,\dots,N$ provided that $C^h_0=C^g_0=0$. The system are
\begin{gather*}
\sum_{j=1}^N C^h_j\int_{\Sigma_j}\frac{\xi^k \dd\xi}{w_+(\xi)} = -\frac{\ii\pi}{2}\int_{\mathbb{R}}\frac{\la^k n(\la)}{w(\la)}\dd\la, \qquad k=0,\dots,N-1.
\end{gather*}

Similarly to \eqref{Cg} this system has a unique solution. In this case the parameter $g_0$ is equal to
\begin{gather}\label{g01}
g_0=-f_0 - \frac{1}{2\pi \ii} \sum_{j=1}^{N}\int_{\Sigma_j}\frac{C^h_j \xi^N}{w_+(\xi)}\dd\xi -
\frac{1}{4}\int_{\mathbb{R}}\frac{\la^N n(\la)}{w(\la)}\dd\la.
\end{gather}
Substituting \eqref{f0} in \eqref{g01} and using equality $C^f_j -C^h_j =-C^g_j $ we obtain
\begin{gather*}
g_0=\sum\limits_{j=0}^N\Re E_j- \frac{1}{2\pi \ii} \sum_{j=1}^{N}\int_{\Sigma_j}\frac{C^g_j \xi^N}{w_+(\xi)}\dd\xi -
\frac{1}{4}\int_{\mathbb{R}}\frac{\la^N n(\la)}{w(\la)}\dd\la,
\end{gather*}
which coincides with \eqref{g0} and thus \eqref{g-hyper} is proved. Besides, we found relations \eqref{g-hyper} and \eqref{g01} between the phase functions $f(z)$ and $g(z)$.

\section[Explicit construction of the matrix $M(t,x,z)$]{Explicit construction of the matrix $\boldsymbol{M(t,x,z)}$}\label{sec6}

In this section we present an explicit construction of $M(t,x,z)$ which solves the RH problem \eqref{RHM}, \eqref{JM}. The main ideas of such a~construction are borrowed in \cite{DIZ97,KMM03, KS}).

First, define
\begin{gather*}
\varkappa(z)=\sqrt[4]{\prod\limits_{j=0}^{N} \frac{z-E^*_j}{z- E_j}},\qquad z\in\mathbb{C}\setminus\Gamma,\qquad \Gamma=\bigcup\limits_{j=0}^N [E_j, E^*_j],
\end{gather*}
where cuts are chosen along $[E_j, E^*_j]$, $j=0,\dots, N$ with orientation from top to bottom. The branch of root is fixed by the condition $\varkappa(\infty)=1$. Then
\begin{gather}\label{kappa}
\varkappa_-(z)=\ii\varkappa_+(z),\qquad z\in \Gamma.
\end{gather}
Notice also that
\begin{itemize}\itemsep=0pt
\item $\varkappa(z)=(z-E_j)^{-1/4}+\ord(1)$ as $z\to E_j$,
\item $\varkappa(z)=1+\sum\limits_{j=0}^{N}\frac{E_j -E^*_j}{4z} +\ord\big(z^{-2}\big)$, $z\to\infty. $
\end{itemize}

Recall that RH problem for $M(z)$ \eqref{RHM}, \eqref{JM}) is as follows:
\begin{itemize}\itemsep=0pt
\item $M(t,x,z)$ is analytic in $\mathbb{C}\setminus\Gamma$, $\Gamma=\bigcup\limits_{j=0}^N [E_j, E^*_j]$;
\item boundary values $M_\pm(t,x,z)$ are continuous except end-points $E_j$ and $E^*_j$ where $M_\pm$ have square integrable singularities;
\item $M_-(t,x,z)=M_+(t,x,z)J_M(t,x,z)$, $z\in\Gamma$,
\begin{align}\label{n0}
 J_M(t,x,z)& =\begin{pmatrix}
 0 &\ii e^{-\ii \phi_0} \\
 \ii e^{\ii \phi_0} & 0
 \end{pmatrix}, \qquad z\in(E_0, E^*_0), \\
& =\begin{pmatrix}0&\ii e^{-\ii xC^f_j-\ii tC^g_j-\ii \phi_j}\\
						\ii e^{\ii xC^f_j+\ii tC^g_j+\ii \phi_j}&0
 \end{pmatrix}, \qquad z\in(E_j, E^*_{j}) \label{n-ge-1}
\end{align}
for $j=1,2,\dots,N$, and $C^f_j$, $ C^g_j$, $\phi_j$ are some given real constants (recall that $C^f_0=C^g_0=0$);
\item $M(t,x,z)=\sigma_2M^*(t,x,z^*)\sigma_2$;
\item $M(t,x,z)=I+\ord\big(z^{-1}\big)$, $z\to\infty$.
\end{itemize}

First, consider the case $N=0$. Then, by \eqref{n0}, $M(t,x,z)\equiv M(z)$ can be constructed using~$\varkappa(z)$
\begin{gather*}
M(z)=\begin{pmatrix}
 \dfrac{\varkappa(z)+\varkappa^{-1}(z)}{2} & \dfrac{\varkappa(z)-\varkappa^{-1}(z)}{2}e^{-\ii \phi_0} \vspace{1mm}\\
 \dfrac{\varkappa(z)-\varkappa^{-1}(z)}{2}e^{\ii \phi_0} & \dfrac{\varkappa(z)+\varkappa^{-1}(z)}{2}
 \end{pmatrix}.
\end{gather*}
Expanding $M(z)$ as $z\to\infty$,
\begin{gather*}
M(z)=I+\frac{m}{z}+\ord\big(z^{-2}\big),
\end{gather*}
we have
\begin{gather*}
m=\frac{ E_0-E^*_0}{4}\begin{pmatrix}
0 &e^{-\ii \phi_0} \\e^{\ii \phi_0} & 0
\end{pmatrix}
\end{gather*}
and thus the simplest periodic solution of the Maxwell--Bloch equation associated with~$\Psi$~(\ref{RHPsi}) has the form of a plane wave (see \eqref{0zonE}--\eqref{0zonN} at the end of the paper).

In order to present an explicit solution of the RH problem in the general case ($N\ge 1$), we introduce necessary facts from the theory of the Riemann manifolds by following closely to \cite{BBEIM,KMM03,KS}. First, let $\mathcal{X}$ be the Riemann surface of genus~$N$ defined by the equation $w^2=P(z)$, where
\begin{gather*}
P(z) = \prod\limits_{j=0}^{N}(z-E_j)(z- E^*_j),
\end{gather*}
with cuts along $\Sigma_j = (E_j, E^*_j)$, $j=0,1,2,\ldots,N$. The Riemann surface $\mathcal{X}$ can be viewed as a~double covering of the complex $z$- plane: two sheets of $z$-plane are glued along $\Sigma_j$. The upper and lower sheets of $\mathcal{X}$ are denoted by~$\mathcal{X}_+$ and~$\mathcal{X}_-$ respectively; they are fixed by the relations
\begin{gather*}
\sqrt{P(z)} =\pm z^{N+1}\big(1+\ord\big(z^{-1}\big)\big), \qquad z=\pi(\mathcal{P})\to \infty, \qquad \mathcal{P}\in {\mathcal X}_\pm,
\end{gather*}
where $z=\pi(\mathcal{P})$ is the standard projection of $\mathcal{P}=(w,z)\in {\mathcal X}$ on the Riemann sphere ${\mathbb{CP}}^1$. Thus each point on the $z$-plane has two preimages $\mathcal{P}_\pm =\mathcal{X}_\pm $, except for the branch points. Denote the preimage of $z=\infty$ on ${\mathcal X}_\pm$ by, respectively, $\infty^\pm$. With the inclusion of two points ($\infty^+, \infty^-$), $\mathcal{X}$ becomes a compact Riemann surface of genus~$N$. The square root $\sqrt{P(z)}$ turns into a meromorphic function on its own compact Riemann surface $\mathcal{X}$, which have $2N+2$ zeros at $E_j$ and $ E^*_j$, $j=0, 1, 2, \dots, N$, and two poles at $\infty^+$ and $ \infty^-$, each of multiplicity $N+1$.

Further, we introduce the Abelian integrals
\begin{gather*}
\omega_j(z)=\int_{E^*_0}^z \psi_j(s)\dd s,\qquad j=1,2,\ldots,N,
\end{gather*}
where $\dd\omega_j(\mathcal P)$ is a basis of holomorphic differentials on~$\mathcal X$
\begin{gather*}
\psi_j(z)=\frac{\sum\limits_{i=1}^{N} c_{ji}z^{N-i}}{\sqrt{P(z)}}.
\end{gather*}
The coefficients $c_{jl}$ are uniquely determined by the normalization conditions
\begin{gather*}
\int_{{\B a}_l} \dd\omega_j(\mathcal{P})=2\int^{E^*_l}_{E_l} \psi_{j+}(z)\dd z=\delta_{jl},\qquad j,l=1,2,\ldots,N.
\end{gather*}

\begin{figure}[ht]\centering
\includegraphics{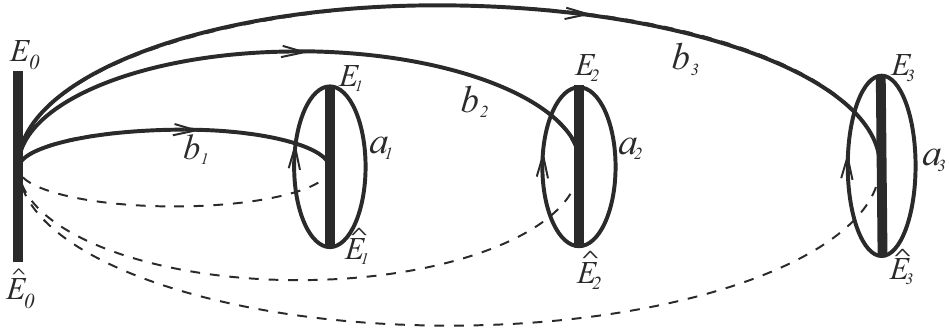}
\caption{$\bf a$- and $\bf b$-cycles.}\label{fig2}
\end{figure}

We have chosen ${\B a}_l$-cycles as ovals on the upper sheet of $\mathcal X$ around the intervals $\big(E_l, \hat E_l\big)$, $\hat E_l:=E_l^*$, $l=0,1,2,\ldots,N$, see Fig.~\ref{fig2}.

The normalized holomorphic differentials define the $b$-period matrix as
\begin{gather*}
B_{jl}=\int_{{\B b}_l} \dd\omega_j(\mathcal{P})=2\sum\limits_{k=1}^{l}\int^{E_{k}}_{E^*_{k-1}} \psi_j(z)\dd z,
\end{gather*}
where ${\B b}_l$-cycle starts from $({E_0}, E^*_{0})$, goes on the upper sheet to $({E_l}, E^*_l)$, and returns on the lower sheet to the starting point. This is a symmetric matrix with positive definite imaginary part.

Let $\B e_j=(0,\ldots,1,\ldots,0)$ be the unit vector in $\mathbb{C}^N$ and $B\B{e}_j$ the $j$-th column of the matrix $B$. Denote by $\Lambda\subset \mathbb{C}^N$ the lattice generated by the linear combinations, with integer coefficients, of the vectors $\B{e}_j$ and $B\B{e}_j$ for $j=1,2, \ldots, N$. Then, by the definition, Jacobian variety of $\mathcal{X}$ is the complex torus $\operatorname{Jac}\{\mathcal{X}\}=\mathbb{C}^N/\Lambda$. The Abel mapping $\B{A}\colon \mathcal{X}\to \operatorname{Jac}\{\mathcal{X}\}$ is defined as follows
\begin{gather}\label{AP}
A_j(\mathcal{P})=\int_{\mathcal{P}_0}^{\mathcal{P}} \dd\omega_j(\mathcal{Q}), \qquad j=1,2, \ldots, N,
\end{gather}
where the point $\mathcal{P}_0 $ is fixed by condition $\pi(\mathcal{P}_0 )=E^*_0$ and $\mathcal{Q}$ is the integration variable. The Abel mapping is also defined for integral divisors $\mathcal{D}=\mathcal{P}_1+\dots+\mathcal{P}_m$ by summation
\begin{gather}\label{AD}
\B{A}(\mathcal{D})=\B{A}(\mathcal{P}_1)+\dots+\B{A}(\mathcal{P}_m)
\end{gather}
and is extended to non-integral divisors $\mathcal{D}=\mathcal{D}^+-\mathcal{D}^-$ (where $\mathcal{D}^\pm$ are integral divisors) by $\B{A}(\mathcal{D})=\B{A}(\mathcal{D}^+)$ $-\B{A}(\mathcal{D}^-)$. If the degree of the divisor $\mathcal{D}$ is zero, then $\B{A}(\mathcal{D})$ is independent of the chosen point $ \mathcal{P}_0$. The Abel theorem states that if $\mathcal{D}=\mathcal{D}^+-\mathcal{D}^-$ is the divisor of a~meromorphic function on the compact Riemann surface $\mathcal{X}$ and $\mathcal{D}^+$, $\mathcal{D}^-$ are integral divisors of zeros and poles, then $\B{A}(\mathcal{D})=0$ in the Jacobian (${\rm mod} \ \Lambda$). Besides, for any non-special integral divisor $\mathcal{D}=\mathcal{P}_1+\dots+ \mathcal{P}_N$ of degree $N$, there exists a vector $\B{w(\mathcal{D})}$ such that the Riemann theta function $\Theta(\B{A}(\mathcal{P})+\B{w(\mathcal{D})})$ defined on $\mathcal{X}$ with cuts along of the cycles $\B a_j $ and $\B b_j$ has precisely~$N$ zeros at $\mathcal{P}_j$, $j=1,\dots,N$. The vector $\B{w(\mathcal{D})}$ is defined by
\begin{gather*}
\B{w}(\mathcal{D})=-\B{A}(\mathcal{D}) -\B{K}.
\end{gather*}
In the hyperelliptic case, the Riemann constant vector $\B{K}$ is defined by (cf.~\cite{Zver})
\begin{gather}\label{K}
K_j=\frac{1}{2}\sum_{\substack{l=1}}^N B_{lj}-\frac{j}{2} \qquad \operatorname{mod} \ \Lambda.
\end{gather}

Associated with the matrix $B$ there is the Riemann theta function defined for $\B{u}\in\mathbb{C}^N$ by the Fourier series
\begin{gather}\label{theta}
\Theta(u_1, \ldots, u_n)=\sum\limits_{\B{l}\in\mathbb{Z}^N}\exp{\{ \pi\ii(B\B{l},\B{l})+2\pi\ii (\B{l},\B{u}) \}},
\end{gather}
where $(\B{l},\B{u})= l_1u_1+\dots+l_Nu_N$. It is an even function, i.e., $\Theta(-{\bf u})=\Theta({\bf u})$, and has the following periodicity properties
\begin{gather*}
\Theta({\bf u}\pm{\bf e}_j)=\Theta({\bf u}), \qquad \Theta({\bf u}\pm B{\bf e_j})=e^{\mp 2\pi\ii u_j-\pi\ii B_{jj}}\Theta(\bf{u}),
\end{gather*}
where $\B e_j=(0, \dots, 0, 1, 0, \dots, 0)$ is the j-th basis vector in $\mathbb{C}^N$. This implies that the function
\begin{gather*}
h({\bf u})= \frac{\Theta({\bf u}+{\bf c}+{\bf d})}{\Theta({\bf u}+{\bf d})},
\end{gather*}
where ${\bf c, \bf d}\in\mathbb{C}^N $ are arbitrary constant vectors, has the periodicity properties
\begin{gather*}
h({\bf u}\pm{\bf e}_j)=h({\bf u}), \qquad h({\bf u}\pm B{\bf e_j})=e^{\mp2\pi\ii c_j }
h({\bf u}).
\end{gather*}

The Abelian integrals ${\bf A}(z)$ considered on the upper sheet of $\mathcal{X}$ ($z\in\mathbb{C}\setminus\Gamma $), have the properties
\begin{gather}\label{Ab-jumps}
{\bf A}_-(z)-{\bf A}_+(z)=0 \qquad \big({\rm mod} \ \mathbb{Z}^N\big),\qquad z\in\mathbb{R}\setminus\cup_{j-0}^N\Re E_j, \\
 {\bf A}_-(z)+{\bf A}_+(z)=0, \qquad z\in (E_0, E_0^*), \\
{\bf A}_-(z)+{\bf A}_+(z)=B{\bf e}_j, \qquad z\in( E_j, E_j^*), \qquad j=1,\dots,N.\label{Ab-jumps3}
\end{gather}
Indeed, since $\psi_l(z_+)+\psi_l(z_-)=0$ on $(E_j, E^*_j)$ and $\psi_l(z)$ is continuous on $\mathbb{C}\setminus\cup_{j=0}^N(E_j, E^*_j) $, it easy to see that for $l=1,2,\ldots,N$,
\begin{gather*}
A_l(z_+) - A_l(z_-)=\int_{E^*_0}^z(\psi_l(s)-\psi_l(s))\dd s=0, \qquad z_\pm\in(-\infty, \Re E_0)\cup(\Re E_{0}, \Re E_1)
\end{gather*}
and
\begin{gather*}
A_l(z_+) - A_l(z_-)=2\sum\limits_{k=1}^{j-1}\int_{E_k}^{E^*_k}\psi_l(s_+)\dd s=\sum\limits_{k=1}^{j-1}\delta_{kl}=0\qquad ({\rm mod} \ \mathbb{Z})
\end{gather*}
for $z_\pm\in(\Re E_{j-1}, \Re E_j)$, $1<j\le N$ and for $z_\pm\in(\Re E_N, +\infty)$. On the other hand,
\begin{gather*}
A_l(z_+) + A_l(z_-)=\int_{E^*_0}^z(\psi_l(s_+)+\psi_l(s_-))\dd s=0,\qquad z_\pm\in(E_0, E^*_{0})
\end{gather*}
and
\begin{gather*}
A_l(z_+) + A_l(z_-)=2\sum\limits_{k=1}^{j}\int^{E_{k}}_{E^*_{k-1}} \psi_j(s)\dd s + \sum\limits_{k=1}^{j-1}\delta_{kl}=B_{jl}, \\
z_\pm\in(E_j, E^*_j), \qquad j,l=1,2,\ldots,N.
\end{gather*}

Now define ($s=1,2$)
\begin{gather}\label{F-H}
F_s(z)= \frac{\Theta({\bf A}(z)+{\bf c}+{\bf d}_s)}{\Theta({\bf A}(z)+{\bf d}_s)},\qquad
H_s(z)= \frac{\Theta(-{\bf A}(z)+{\bf c}+{\bf d}_s)}{\Theta(-{\bf A}(z)+{\bf d}_s)},\qquad z\in\mathbb{C}\setminus\Gamma,
\end{gather}
where ${\bf c, {\bf d}_1, {\bf d}_2}\in\mathbb{C}^N$ are arbitrary (so far) constant vectors. Then, by \eqref{Ab-jumps}--\eqref{Ab-jumps3}, we have ($s=1,2$)
\begin{gather*}
F_{s-}(z)=F_{s+}(z), \qquad H_{s-}(z)=H_{s+}(z), \qquad z\in\mathbb{R}\setminus\bigcup\limits_{j=0}^N \Re E_j
\end{gather*}
and
\begin{gather*}
F_{s-}(z)=e^{-2\pi\ii c_j } H_{s+}(z), \qquad H_{s-}(z)= e^{2\pi\ii c_j }F_{s+}(z), \qquad z\in (E_j, E^*_j)
\end{gather*}
for $j=0,1,\ldots,N$, where $c_0=0$.

Next, define the matrix-valued function
\begin{gather*}
\hat M(z):=\begin{pmatrix}
 F_1(z)& H_1(z) \\
 F_2(z) & H_2(z)
 \end{pmatrix}.
\end{gather*}
Then we have
\begin{subequations} \label{N-jumps}
\begin{gather}
 \hat M_-(z)=\hat M_+(z), \qquad z\in\mathbb{R}\setminus\bigcup\limits_{j=0}^N \Re E_j,\\
 \hat M_-(z)=\hat M_+(z)\begin{pmatrix}
 0& e^{2\pi\ii c_j}\\
 e^{-2\pi\ii c_j}&0
 \end{pmatrix}, \qquad z\in( E_j, E^*_j),\qquad j=0,1,\ldots,N,
\end{gather}
\end{subequations}
with $c_0=0$. Finally, taking into account (\ref{kappa}) and (\ref{N-jumps}), we define $M$ by (provided $F_1(\infty)\ne 0$ and $H_2(\infty)\ne 0$)
\begin{gather}	\label{M-theta}
M(z):= \begin{pmatrix}
 a(z) \dfrac{F_1(z)}{F_1(\infty)} &b(z) \dfrac{H_1(z)}{F_1(\infty)}e^{-\ii\phi_0} \vspace{1mm}\\
 b(z) \dfrac{F_2(z)}{H_2(\infty)}e^{\ii\phi_0} &a(z) \dfrac{H_2(z)}{H_2(\infty)}
 \end{pmatrix},
\end{gather}
where $a(z):= \frac{1}{2}(\varkappa(z)+\varkappa^{-1}(z))$ and $b(z):= \frac{1}{2}\big(\varkappa(z)-\varkappa^{-1}(z)\big)$. Obviously, $M(z)$ is analytic in $\mathbb{C}\setminus\bigcup\limits_{j=0}^N [E_j, E^*_j]$,
$M(z) = I+\ord\big(z^{-1}\big)$ as $z\to\infty$, and, due to~(\ref{N-jumps}), it has the jumps
\begin{gather*}
M_-(z)= M_+(z)\begin{pmatrix}
 0 & \ii e^{2\pi\ii c_j}e^{-\ii\phi_0}\\
 \ii e^{-2\pi\ii c_j}e^{\ii\phi_0} & 0
 \end{pmatrix}, \qquad z\in(E_j, E^*_j),\qquad j=0,1,\ldots,N,
\end{gather*}
if we take into account that $a_-(z)=\ii b_+(z)$ and $b_-(z)=\ii a_+(z)$ when $z\in( E_j, E^*_j)$, $j=0,1,\ldots,N$. These jumps are consistent with the jump conditions (\ref{n0}), (\ref{n-ge-1}) for the RH problem, if we
set ${c_j}:= - \frac{t{C^f_j} +x{C^g_j}+{\phi_j}}{2\pi}$, $j=1,\dots, N$ (recall that $c_0=0$) and hence the matrix $M(z)$ depends from $t,x\in\mathbb{R}$ additionally, i.e., $M(z)=M(t,x,z)$.

It remains to choose the vectors $\B d_1$ and $\B d_2$ in such a way that $M(z)$ is analytic at the zeros of the denominators in (\ref{F-H}), i.e., the zeros of $\Theta({\bf A}(z)+{\bf d}_s)$ and $\Theta(-{\bf A}(z)+{\bf d}_s)$ ($s=1,2$) are to be canceled by the zeros of $\varkappa(z)\pm\varkappa^{-1}(z)$.

The zeros of $\varkappa(z)\pm \varkappa^{-1}(z)$ are those of $\varkappa^2(z)\pm 1$, and hence of $\varkappa^4(z) - 1$. By the definition of $\varkappa(z)$, equation $\varkappa^4(z) - 1 = 0$ reads $r(z)=1$, where
\begin{gather*}
r(z) :=\prod\limits^{N}_{j=0} \frac{z-E^*_j}{z- E_j}.
\end{gather*}
Since $\sum\limits^{N}_{j=0}(E_j-E^*_j)=2\ii\Im E_j\neq0$, equation $r(z)=1$ reduces to
\begin{gather*}
0=\prod\limits^{N}_{j=0} {(z-E^*_j)} -\prod\limits^{N}_{j=0} {(z-E_j)}=2\ii \sum^{N}_{j=0}\Im E_j\prod\limits^{N-1}_{l=0}(z-z_l)
\end{gather*}
with some finite $z_l$, $l=0,\dots, N-1$.

Introduce the non-special divisor $\mathcal{D}=\mathcal{P}_1+\dots+\mathcal{P}_N$ such that $\mathcal{D}=\mathcal{D}_1+\mathcal{D}_2$, where $\mathcal{D}_1 =\mathcal{P}_1+\dots+\mathcal{P}_{N_1}\in \mathcal{X}_-$, $0\le N_1\le N$ and $a(z_j)=\varkappa(z_j)+\varkappa^{-1}(z_j)=0$, $z_j=\pi(\mathcal{P}_{j})$, $j=1, 2, \ldots, N_1$, whereas $\mathcal{D}_2 =\mathcal{P}_{N_1+1}+\dots+\mathcal{P}_N\in \mathcal{X}_+$ with $b(z_j)=\varkappa(z_j)-\varkappa^{-1}(z_j)=0$, $j=N_1+1, N_1+2, \ldots, N$. Set $\B d_1=\B A(\mathcal{D})+\B K$ and $\B d_2=-\B A(\mathcal{D})-\B K$. Then $\Theta(\B A(\mathcal{P})+\B A(\mathcal{D})+\B K)$ has $N_1$ zeroes
$\mathcal{P}^\prime_1,\ldots,\mathcal{P}^\prime_{N_1}$ on $\mathcal{X}_+$ and $N-N_1$ zeroes $\mathcal{P}^\prime_{N_1+1},\ldots,\mathcal{P}^\prime_N$ on $\mathcal{X}_-$~\cite{BBEIM}, where the points $\mathcal{P}^\prime_j$ and $\mathcal{P}_j$ form a conjugated pair of points on~$\mathcal{X}$ with $\pi(\mathcal{P}^\prime_j)=\pi(\mathcal{P}_j)=z_j\in\mathbb{C}$, $j=1,2,\ldots,N$. Similarly, $\Theta(\B A(\mathcal{P})-\B A(\mathcal{D})-\B K)$ has $N-N_1$ zeroes $\mathcal{P}_{N_1+1}, \ldots, \mathcal{P}_N$ on $\mathcal{X}_+$ and~$m$ zeroes $\mathcal{P}_1,\ldots,\mathcal{P}_{N_1}$ on~$\mathcal{X}_-$. Taking the restrictions of these Riemann theta functions on the upper sheet with the cut $\Gamma=\bigcup\limits_{j=0}^N (E_j, E^*_j)$, we have that
\begin{gather*}
F_1(z)= \frac{\Theta({\bf A}(z)+{\bf c}+{\bf d}_1)}{\Theta({\bf A}(z)+{\bf d}_1)}, \qquad H_2(z)= \frac{\Theta(-{\bf A}(z)+{\bf c}+{\bf d}_2)}{\Theta(-{\bf A}(z)+{\bf d}_2)}
\end{gather*} are analytic in $z\in\mathbb{C}\setminus\Gamma $ with poles at $z_1,\ldots, z_{N_1}$, which are canceled in the products $a(z)F_1(z)$ and $a(z)H_2(z)$. Similarly, $b(z)F_2(z)$ and $b(z)H_1(z)$ are analytic in $\mathbb{C}\setminus\Gamma$, since the poles $z_{N_1+1}, \ldots, z_N$ are canceled by the zeroes of~$b(z)$. Notice that the idea to cancel the poles of~$F_j$ and~$H_j$ by the zeros of $\varkappa\pm\varkappa^{-1}$ goes back to \cite{DIZ97,DVZ94,DVZ97}.

Thus matrix $M(z)$ \eqref{M-theta} satisfies all conditions to be a solution of the RH problem \eqref{RHM}--\eqref{JM} if only
\begin{gather*}
F_1(\infty)=\dfrac{\Theta({\bf A}(\infty^+)+\B A(\mathcal{D})+\B K+{\bf c})} {\Theta({\bf A}(\infty^+)+\B A(\mathcal{D})+\B K)}, \qquad
H_2(\infty)=\dfrac{\Theta({\bf A}(\infty^+)+\B A(\mathcal{D})+\B K-{\bf c})} {\Theta({\bf A}(\infty^+)+\B A(\mathcal{D})+\B K)}
\end{gather*}
do not vanish. Since the divisor $\mathcal{D}$ is non-special and $\infty^\pm\notin\mathcal{D}$, the denominator does not equal zero and takes a finite value. It is well known (cf.~\cite{BBEIM}) that the divisor of zeroes of $\Theta({\bf A}(z)+\B A(\mathcal{D})+\B K\pm{\bf c})$ remain non-special if vector $\bf c$ is sufficiently small. Since all zeroes of $F_1(z)$ ($H_2(z)$) belong to the mentioned divisor and this divisor does not contain infinity, then $F_1(\infty)\neq0$ (as well as $H_2(\infty)\neq0$). Moreover, taking into account the symmetries $M(z)=\sigma_2M^*(z^*)\sigma_2$, i.e., $M_{22}(z)=M_{11}^*(z^*)$ and $M_{21}(z)=-M_{12}^*(z^*)$, and unity determinant of the matrix $M$ we obtain
\begin{gather}\label{boundM}
\big(M_{11}(z)M_{22}(z)-M_{12}(z)M_{21}(z)\big)\vert_{z=\la}= |M_{11}(\la)|^2+|M_{12}(\la)|^2\equiv 1, \qquad \la\neq\Re E_j
\end{gather}
that gives the boundedness of all entries of matrix $M(\la)$. In turn, it means that $F_1(\infty)$ and~$H_2(\infty)$ can not vanish for any vector $\mathbf c$. To prove the symmetry of $M(z)$ let us consider matrix $\tilde M(z):=\sigma_2M^*(z^*)\sigma_2$. Then $\tilde M(z)$ and the original matrix $M(z)$ solve the same RH problem and, due to the uniqueness of the solution of the RH problem, we have $\tilde M(z)\equiv M(z)$. Hence $M(z)$ satisfies the symmetry conditions.

We have constructed the matrix $M(z)=M(t,x,z)$ that solves the required RH problem \eqref{RHM}--\eqref{JM} and thus $M$ provides analyticity of $\Psi(t,x,z)$ in $z\in\mathbb{C}\setminus\Sigma$, and also continuity up to the contour~$\Sigma$ (except for the endpoints $E_j$ and $E^*_j$, where $\Psi$ has weak singularities).

Formulas for entries $M_{ij}(t,x,z)$ of the matrix $M(t,x,z)$ follows from \eqref{F-H} and \eqref{M-theta}. Expanding it at infinity,
\begin{gather*}
M(t,x,z)=I+\frac{m(t,x)}{z}+\ord\big(z^{-2}\big), \qquad z\to\infty,
\end{gather*}
and taking into account that
 $\Theta(\B A(z)+\B B)$ is bounded and $ \frac{\dd\B A}{\dd z}=\ord\big(z^{-2}\big)$ as $z\to\infty$, we have
\begin{gather}
m_{12}(t,x)=E_0e^{-\ii\phi_0}\frac{\Theta(-\B A(\infty)+\B A(\mathcal{D})+\B K+\B C(t,x))\Theta(\B A(\infty)+\B A(\mathcal{D})+\B K)}{\Theta(-\B A(\infty)+\B A(\mathcal{D})+\B K)\Theta(\B A(\infty)+\B A(\mathcal{D})+\B K+\B C(t,x))},\label{m12m21}\\
m_{21}(t,x)=E_0e^{\ii\phi_0}\frac{\Theta(\B A(\infty)-\B A(\mathcal{D})-\B K+\B C(t,x))\Theta(\B A(\infty)+\B A(\mathcal{D})+
\B K)}{\Theta(\B A(\infty)-\B A(\mathcal{D})-\B K)\Theta(\B A(\infty)+\B A(\mathcal{D})+\B K-\B C(t,x))},\nonumber
\end{gather}
where $E_0= \frac{1}{4}\sum\limits^{N}_{j=0}(E_j-E^*_j)= \frac{\ii}{2}\sum\limits^{N}_{j=0}\Im E_j$ and $\B C(t,x):=- \frac{t{\B C^f} +x{\B C^g}+{\bf \phi}}{2\pi}$. Notice that~$\B C^f$ and~$\B C^g$ are determined when constructing $f(z)$ and $g(z)$ whereas $E_j=\Re E_j+\ii\Im E_j$, $j=0,1,2,\ldots, N$, and real constants $(\phi_0, \phi_1, \phi_2, \ldots, \phi_N)$ present itself free real parameters total number of which is equal to $3N+3$. Evidently, the constant $\phi_0$ is defined modulo $2\pi$, while $(2\pi\phi_1, 2\pi\phi_2, \ldots, 2\pi\phi_N)$ can be regarded as a vector on the Jacobian $\operatorname{Jac}\{\mathcal{X}\}$. Formulas~\eqref{m12m21} will be used for a definition of finite-gap solutions to the MB equations.

In the theory of finite-gap integration \cite{BBEIM}, the divisor $\mathcal{D}$ is taken to be arbitrary, it defines poles of the Baker--Akhiezer vector function. In the absence of symmetry, such a vector function satisfies corresponding AKNS equations defined by two complex valued functions. These equations generate the focusing NLS equation for unique complex valued function if and only if they possess a symmetry which, in turn, take place if and only if the so called reality conditions
\begin{gather}
\prod\limits_{j=0}^{N}(z- E_j)(z-E^*_j) -|q(0,0)|^2\prod\limits_{j=1}^{N}(z-z_j)(z-z^*_j)\nonumber\\
\qquad{} =\big(z^{N+1}+f_{N}z^{N}+f_{N-1}z^{N-1}+\dots +f_0\big)^2\label{RC}
\end{gather}
are fulfilled. The left hand side of the equality is determined by $4N+3$ real parameters: branching points $E_j\in\mathbb{C}$ ($\Im E_j\neq 0$, $j=0, 1, \ldots, N$), projections $z_j=\pi(\mathcal{P}_j)$ of the non-special divisor $\mathcal{D}=\mathcal{P}_1+\dots+\mathcal{P}_N$, and $q(0,0)$ where $q(x,t)$ is a finite-gap solution of NLSE. The reality conditions reads as follows: the difference of the polynomials on the left-hand side must be a square of some polynomial $N + 1 $-th degree with real coefficients. They contain $N$ nonlinear relations (because independently from $z_j$ $f_N=-\sum\limits_{j=0}^N \Re E_j$) and hence the number of independent real parameters decreases to $3N+3$. This conditions were first obtained in~\cite{K76} (see also~\cite{IK76}). The same conditions \eqref{RC} characterize a set of finite-gap Dirac operators with anti-Hermitian potential matrices~\cite{K76}. Therefore conditions \eqref{RC} are also applicable to our case where the total number of free (real) parameters is also equals to $3N+3$. In our case the divisor $\mathcal{D}$ is fixed by zeroes $z_j$ of the functions $\varkappa(z)\pm\varkappa^{-1}(z)$. Thus, the reality conditions mean that there is a correspondence between ($\phi_0, \phi_1, \ldots, \phi_N$) and ($f_0, f_1, \ldots, f_{N-1}, |q(0,0)|=|{\mathcal E}(0,0)|$). However, this issue is beyond the scope of article. In our approach $E_0, E_1, \ldots, E_N$ and $\phi_0, \phi_1, \ldots, \phi_N$ are independent. Due to the symmetry $\Psi(t, x, z)=\sigma_2\Psi^*(t, x, z^*)\sigma_2$ the potential matrix $H(t,x)$ is anti-Hermitian for any choice of the parameters (see the next section for details).

\section[AKNS equations for $\Psi(t,x,z)$]{AKNS equations for $\boldsymbol{\Psi(t,x,z)}$}\label{sec7}

Here we prove Theorem \ref{PsiAKNS}.

\begin{proof}By the construction, $\Psi$ is analytic with respect to $\bf c=\bf C(t,x)$ which, in turn, is linear with respect to $t$ and $x$. Hence $\Psi(t,x,z)$ is smooth in $t,x\in\mathbb{R}$. The matrix $\Psi(t,x,z)$ is also analytic in $z\in\mathbb{C}\setminus\Sigma$ and has (due to \eqref{RHM} and \eqref{JM}) the jump across $\Sigma$
\begin{gather*}
\Psi_-(t,x,z)=\Psi_+(t,x,z) J(x,z),
\end{gather*}
where the jump matrices
\begin{align*}
\qquad J(x,z)&=\begin{cases}\begin{pmatrix}
 \ee^{-\frac{\pi xn(\la)}{2}} & 0 \\
 0 & \ee^{\frac{\pi xn(\la)}{2}}
 \end{pmatrix}, \qquad\ \la\in\mathbb{R}\setminus\bigcup\limits_{j=0}^N\Re E_j,\\
 \begin{pmatrix}
 0 & \ii e^{-\ii \phi_j}\\
 \ii e^{\ii \phi_j} & 0
 \end{pmatrix}, \quad z\in\Sigma_j=(E_j, E^*_j), \qquad j=0,1,\dots,N,
\end{cases}
\end{align*}
are independent on $t$. The jump condition gives
\begin{gather*}
\frac{\partial\Psi_-(t,x,z)}{\partial t}\Psi^{-1}_-(t,x,z)=\frac{\partial\Psi_+(t,x,z)}{\partial t}\Psi^{-1}_+(t,x,z), \qquad z\in\Sigma.
\end{gather*}
This relation, together with a continuity of $\Psi_\pm$ outside of exceptional points ($E_j, E_j^*, \Re E_j$), implies that logarithmic derivative $\Psi_t(t,x,z)\Psi^{-1}(t,x,z)$ is analytic (entire) in $z\in\mathbb{C}$.
Indeed, since $\Psi_t(t,x,z)\Psi^{-1}(t,x,z)$ has no jump across $\Sigma\setminus\bigcup\limits_{j=0}^N\Re E_j$ then it can be extended to a continuous function because the exceptional points are removable singularities. We took into account the boundedness at the points of self intersection $\Re E_j$, weak singularities at the endpoints~$E_j$,~$E_j^*$ and the second order of $\Psi$. Further, since $M(t,x,z)$ and $M_t(t,x,z)$ have the asymptotics:
\begin{gather*}
M(t,x,z)=I+\frac{m(t,x)}{z} +\ord\big(z^{-2}\big),\qquad \frac{\dd M(t,x,z)}{\dd t}=\frac{\dd m(t,x)/\dd t}{z}+O\big(z^{-2}\big), \!\qquad z\in\mathbb{C}_\pm,
\end{gather*}
as $ z\to\infty$, it follows that
\begin{gather*}
\Psi_t(t,x,z)\Psi^{-1}(t,x,z)\\
\qquad{} =-\ii z\sigma_3+\ii \ee^{\ii(tf_0+xg_0)\sigma_3}[\sigma_3, m(t,x)]\ee^{-\ii(tf_0+xg_0)\sigma_3}+\ord\big(z^{-1}\big),\qquad z\to\infty,
\end{gather*}
where $[A, B]:=AB-BA$. Therefore, by Liouville's theorem, the logarithmic derivative is a~polynomial
\begin{gather*}
U(z):=\Psi_t(t,x,z)\Psi^{-1}(t,x,z)=-\ii z\sigma_3-H(t,x),
\end{gather*}
where
\begin{gather*}
H(t,x):=-\ii\ee^{\ii(tf_0+xg_0)\sigma_3}[\sigma_3, m(t,x)]\ee^{-\ii(tf_0+xg_0)\sigma_3}=
\begin{pmatrix}0&q(t,x)\\p(t,x)&0\end{pmatrix}.
\end{gather*} Using the symmetry $\sigma_2 \Psi ^*(z^*)\sigma_2=\Psi(t,x,z)$ we find that $U(z)=\sigma_2 U^*(z^*)\sigma_2$. This symmetry implies that $H$ is anti-Hermitian, i.e., $H= -H^\dag $. Hence $q(t,x)=-p^*(t,x)$ and we put $q(t,x):={\mathcal E}(t,x)/2$ where ${\mathcal E}(t,x)=-4\ii m_{12}(t,x)\ee^{2\ii(tf_0+xg_0)}$ with $m_{12}(t,x)$ defined in~\eqref{m12m21}. Thus $\Psi(t,x,z)$ satisfies the first equation of~\eqref{PsitPsix} with matrix~$H$ given by~\eqref{H}.

In contrast with previous case logarithmic derivative $\Psi_x(t,x,z)\Psi^{-1}(t,x,z)$ is analytic in $z\in\mathbb{C}_\pm$ only. Indeed, since jump matrix~$J(z)$~\eqref{ephij} is independent on $t$ and $x$ for $z\in\Sigma\setminus\mathbb{R}$, then this logarithmic derivative is continuous across the contour $\Sigma\setminus\mathbb{R}$, while it is not continuous across the real line because the corresponding jump matrix~\eqref{RHPsi} $J(x,\la)=\ee^{-\frac{\pi n(\la) x\sigma_3}{2}}$ depends on $x$. The endpoints of the contour $\Sigma$ are removable singularities by the same reasons as above. Further, the asymptotic behavior at infinity gives
\begin{gather*}
 \Psi_x(t,x,z)\Psi^{-1}(t,x,z)=\ii z\sigma_3 +H(t,x) +O\big(z^{-1}\big), \qquad z\in\mathbb{C}_\pm, \qquad z\to\infty.
\end{gather*}
The jump condition $\Psi_-=\Psi_+ e^{-\frac{\pi x n(\la)\sigma_3}{2}}$ ($\la$ is real) yields
\begin{gather*}
\Psi_x(t,x,\la+\ii0)\Psi^{-1}(t,x,\la+\ii0)-
\Psi_x(t,x,\la-\ii0)\Psi^{-1}(t,x,\la-\ii0)= \frac{\pi n(\la)}{2}F(t,x,\la),
\end{gather*}
where
\begin{gather*}
F(t,x,\lambda):=\Psi(t,x,\la+\ii0)\sigma_3\Psi^{-1}(t,x,\la+\ii0)=\Psi(t,x,\la-\ii0)\sigma_3\Psi^{-1}(t,x,\la-\ii0)\\
\hphantom{F(t,x,\lambda)}{} =\ee^{\ii(tf_0+xg_0)\sigma_3}M(t,x,\la)\sigma_3M^{-1}(t,x,\la)\ee^{-\ii(tf_0+xg_0)\sigma_3},\\
 \la\neq\Re E_j, \qquad j= 0, 1, 2, \dots, N.
\end{gather*}

Therefore $ \Psi_x(t,x,z)\Psi^{-1}(t,x,z)-\ii z\sigma_3 -H(t,x)$ is represented through Cauchy integral
\begin{gather*}
\Psi_x(t,x,z)\Psi^{-1}(t,x,z)-\ii z\sigma_3 -H(t,x)= \frac{1}{4\ii}\int_{-\infty}^\infty \frac{F(t,x,s)n(s)}{s-z}\dd s, \qquad z\notin\mathbb{R}.
\end{gather*}

Due to the symmetries of $M(t,x,\la)$ we find that $F(t,x,\la)$ is Hermitian. Since $\operatorname{tr} (\Psi_x(t,x,\la\pm\ii0)\Psi^{-1}(t,x,\la\pm\ii0)) =(\det\Psi(t,x,\la\pm\ii0))^\prime_x\equiv0$ and $\operatorname{tr}\sigma_3=\operatorname{tr} H(t,x)=0$ then $\operatorname{tr}F(t,x,\la)=0$ and, hence, $F(t,x,\la)$ has the structure
\begin{gather*}
F(t,x,\la):=\begin{pmatrix}{\mathcal N}(t,x,\la)&{\mathcal\rho}(t,x,\la)\\
{\mathcal\rho}^*(t,x,\la)&-{\mathcal N}(t,x,\la)\end{pmatrix}.
\end{gather*}

Thus $\Psi(t,x,z)$ satisfies two differential equations
\begin{gather*}
\Psi_t =U(t,x,z)\Psi, \qquad U(t,x,z)=-\ii z\sigma_3-H(t,x), \\
\Psi_x =V(t,x,z)\Psi, \qquad V(t,x,z)=\ii z\sigma_3+H(t,x)-\ii G(t,x,z),
\end{gather*}
where
\begin{gather*}
G(t,x,z)= \frac{1}{4}\int_{-\infty}^\infty \frac{F(t,x,s)n(s)}{s-z}\dd s, \qquad z\notin\mathbb{R}.
\end{gather*}
For real $z=\la\in\mathbb{R}$ we have two differential in $x$ equations
\begin{gather*}
\Psi_x =V^\pm(t,x,\la)\Psi, \qquad V^\pm(t,x,\la)=\ii\la\sigma_3+H(t,x)- \ii G^\pm(t,x,\la),
\end{gather*}
where $G^\pm(t,x,\la):=G(t,x,\la\pm\ii0)$.

The compatibility condition $(\Psi_{xt}(t,x,\la\pm\ii0)=\Psi_{tx}(t,x,\la\pm\ii0))$ gives the identity in~$\la$
\begin{gather*}
U_x(t,x,\la)-V^\pm_t(t,x,\la)+\big[U(t,x,\la),V^\pm(t,x,\la)\big]=0.
\end{gather*}
This identity is equivalent to
\begin{gather*}
H_t(t,x)+H_x(t,x)- \frac{1}{4}\int_{-\infty}^\infty [\sigma_3,F(t,x,s)]n(s)\dd s \\
\qquad{} =
 \frac{\ii}{4}\int_{-\infty}^\infty \frac{F_t(t,x,s)+[\ii s\sigma_3+H(t,x), F(t,x,s)]}{s-\la\mp\ii0}n(s)\dd s
\end{gather*}
and it is possible if and only if the left and right hand sides are equal zero, i.e.,
\begin{gather*}
H_t(t,x)+H_x(t,x)- \frac{1}{4}\int_{-\infty}^\infty [\sigma_3,F(t,x,s)]n(s)\dd s=0,\\
F_t(t,x,\la)+[\ii\la\sigma_3+H(t,x),F(t,x,\la)] =0.
\end{gather*}
These matrix equations are equivalent to the MB equations (\ref{MB1})--(\ref{MB3}). Thus we proved that the matrices $\Psi(t,x,\la\pm\ii0)$ satisfy equations \eqref{PsitPsix} (with coefficients \eqref{H}, \eqref{FviaM}) which coincide with AKNS system \eqref{zt} and \eqref{zx}. Hence scalar functions ${\mathcal E}(t,x)$, ${\mathcal N}(t,x,\la)$ and ${\mathcal\rho}(t,x,\la)$ satisfy the Maxwell--Bloch equations \eqref{MB1}--\eqref{MB3}.
\end{proof}

\section{Finite-gap solutions to the MB equations}\label{sec8}

Here we prove the Theorem \ref{fgsol}.
\begin{proof}Taking into account \eqref{m12m21} we have
\begin{gather}\label{calE3}
{\mathcal E}(t,x)=E_{\Theta}\ee^{-\ii\phi_0} \frac{\Theta(-\B A(\infty)+\B A(\mathcal{D})+\B K+\B C(t,x))} {\Theta(\B A(\infty)+\B A(\mathcal{D})+\B K+\B C(t,x))}\ee^{2\ii(tf_0+xg_0)},
\end{gather}
where
\begin{gather*}
E_{\Theta}:=2\frac{\Theta(\B A(\infty)+\B A(\mathcal{D})+\B K)}{\Theta(-\B A(\infty)+\B A(\mathcal{D})+\B K)} \sum^{N}_{j=0}\Im E_j
\end{gather*}
is a constant. Hence \eqref{calE3} gives \eqref{solMBE}. Equation \eqref{solF} follows from \eqref{FviaM}. Formulas for entries $M_{ij}(t,x,z)$ of the matrix $M(t,x,z)$ follows from \eqref{F-H} and \eqref{M-theta}. Finally, the analytical dependence of all ingredients of the construction with respect to $\mathbf c=\mathbf C(t,x)$ provides the smoothness of the solution of the MB equations. The important property~\eqref{boundM} follows from~\eqref{H} and a~chain of equalities
\begin{gather*}
\mathcal N^2(t,x,\la)) + |\rho(t,x,\la)|^2=-\det F(t,x,\la)=-\det M(t,x,\la)\sigma_3M^{-1}(t,x,\la)=1.
\end{gather*}
Thus $\mathcal N(t,x,\la))$ and $\rho(t,x,\la)$ are smooth for all $t,x\in\mathbb{R}$ ($\la\neq\Re E_j$) and bounded for all $t,x,\la\in\mathbb{R}$.
\end{proof}

In particulary, the simplest periodic solution to the MB equations takes the form of a plane wave
\begin{gather}\label{0zonE}
{\mathcal E}(t,x)=2\Im E_0\ee^{2\ii(tf_0+xg_0)-\ii\phi_0},\\ \label{0zonrho}
\rho(t,x,\la)=- \frac{\ii\Im E_0}{w(\la)}\ee^{2\ii(tf_0+xg_0)-\ii\phi_0},\\ \label{0zonN}
{\mathcal N}(t,x,\la)= \frac{\la-\Re E_0}{w(\la)},
\end{gather}
where $w(\la)=\sqrt{(\la-E_0)(\la-E^*_0)}$, $\la\in\mathbb{R}$ and
\begin{gather*}
tf_0+xg_0= (x-t)\Re E_0-\frac{x}{4}\int_{-\infty}^\infty\frac{n(\la) \dd\la}{w(\la)}.
\end{gather*}
Some periodic and rational solutions of the reduced Maxwell--Bloch equations with $n(\la)=\delta(\la)$ were recently obtained in~\cite{WWG}.

\section{Final remarks}\label{sec9}

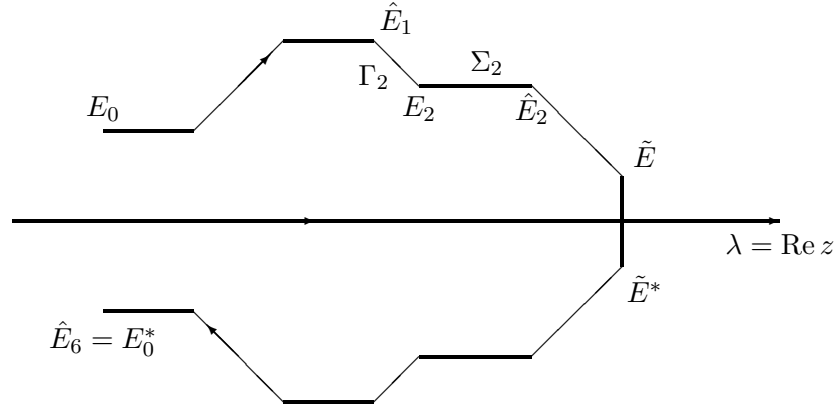
\begin{figure}[ht]
\begin{picture}(150,160)(-70,100)
\setlength{\unitlength}{0.60mm}
\linethickness{1,20pt}
\put(10.00,100.00){\vector(1,0){170.00}}
\linethickness{1,50pt}\put(70.00,100.00){\vector(1,0){7.00}}
\put(30.00,120.00){\line(1,0){20.00}}
\put(50.00,120.00){\line(1,1){20.00}}\put(60.00,130.00){\vector(1,1){7.00}}\put(60.00,70.00){\vector(-1,1){7.00}}
\put(70.00,140.00){\line(1,0){20.00}}
\put(90.00,140.00){\line(1,-1){10.00}}
\put(100.00,130.00){\line(1,0){25.00}}
\put(125.00,130.00){\line(1,-1){20.00}}
\put(30.00,80.00){\line(1,0){20.00}}
\put(50.00,80.00){\line(1,-1){20.00}}
\put(70.00,60.00){\line(1,0){20.00}}
\put(90.00,60.00){\line(1,1){10.00}}
\put(100.00,70.00){\line(1,0){25.00}}
\put(125.00,70.00){\line(1,1){20.00}}
\put(145.00,90.00){\line(0,1){20.00}}
\put(30.00,125.00){\makebox(0,0)[cc]{$E_0$}}\put(30.00,74.00){\makebox(0,0)[cc]{$\hat E_6=E^*_0$}}
\put(150.00,115.00){\makebox(0,0)[cc]{$\tilde E$}}\put(150.00,85.00){\makebox(0,0)[cc]{$\tilde E^*$}}
\put(115.00,135.00){\makebox(0,0)[cc]{$\Sigma_2$}}
\put(90.00,132.00){\makebox(0,0)[cc]{$\Gamma_2$}}
\put(95.00,145.00){\makebox(0,0)[cc]{$\hat E_1$}}
\put(100.00,125.00){\makebox(0,0)[cc]{$E_2$}}
\put(125.00,125.00){\makebox(0,0)[cc]{$\hat E_2$}}
\put(180.00,95.00){\makebox(0,0)[cc]{$\lambda=\Re z$}}
\end{picture}
\caption{Oriented contour $\Sigma$.}\label{fig3}
\end{figure}

Many asymptotic problems deal with contours of another structure and, consequently, another RH problems. Namely, let $\Sigma=\mathbb{R}\cup\bigcup\limits_{j=0}^N\Sigma_j\cup\bigcup\limits_{j=1}^N\Gamma_j$ where $\Sigma_j=(E_j, \hat E_j)$, $j=0,1,\ldots, N$, and $\Gamma_j=(\hat E_{j-1}, E_j)$, $j=1,\ldots, N$ (Fig.~\ref{fig3}). $\Sigma$ has to be symmetric with respect to the real line, therefore we suppose that $E_j^*=\hat E_{N-j}$ and $\hat E_j^*=E_{N-j}$. Denote through $\Re\tilde E$ a unique point of self-intersection of contour $\Sigma$. Here $\tilde E=E_{N/2}$ for even $N$ and $\tilde E=E_{[N/2]+1}$ for odd $N$ where $[N/2]$ is the integer part of~$N/2$.

\begin{Def}Let a contour $\Sigma$, a set of real constants $(\phi_0, \phi_1,\dots,\phi_N)$ and a weight function $n(\la)$ be given. A $2\times2$ matrix $\Psi(t,x,z)$ is called the Baker--Akhiezer function associated with the Maxwell--Bloch equations if for any $x,t\in\mathbb{R}$:
\begin{itemize}\itemsep=0pt
\item $\Psi(t,x,z)$ is analytic in $z\in\mathbb{C}\setminus\overline\Sigma$ where $\overline\Sigma$ is a closure of $\Sigma$;
\item boundary values $\Psi_\pm(t,x,z)$ are continuous with exception of endpoints $E_j$ and $E^*_j$, $j=0,1,\dots,N$ where they have square integrable singularities;
\item $\Psi(t,x,z)$ satisfies the jump conditions:
\begin{gather*}
\Psi_-(t,x,z)=\Psi_+(t,x,z)J(x,z),\qquad z\in\Sigma,
\end{gather*}
where
\begin{gather}
 J(x,z)=\begin{pmatrix}
 \ee^{-\frac{\pi xn(\la)}{2}}& 0 \\
 0 & \ee^{\frac{\pi xn(\la)}{2}}
\end{pmatrix}, \qquad z=\la\in\mathbb{R} \setminus\Re\tilde E,\nonumber\\
 J(x,z)=\begin{pmatrix}
 0 & \ii e^{-\ii \phi_0}\\
 \ii e^{\ii \phi_0} & 0
 \end{pmatrix}, \qquad z\in\Sigma_j=\big(E_j, \hat E_j\big), \qquad j=0,1,\dots,N,\label{ephij}\\
 J(x,z) =\begin{pmatrix}
 e^{-\ii\phi_j}&0\\
 0&e^{\ii\hat\phi_j}
 \end{pmatrix}, \qquad z\in\hat\Gamma_j=\big(\hat E_{j-1}, E_j\big), \qquad j=1,\dots,N,\nonumber
\end{gather}
\item $\Psi(t,x,z)$ satisfies the symmetry condition $\Psi(t,x,z)=\sigma_2\Psi^*(t,x,z^*)\sigma_2$, where
$\sigma_2=\left(\begin{smallmatrix}
 0 & -\ii \\
 \ii & 0 \\
 \end{smallmatrix}\right)$;
\item $\Psi(t,x,z)=\big(I+O\big(z^{-1}\big)\big)e^{-\ii z(t-x)\sigma_3}$ as $z\to\infty$.
\end{itemize}
\end{Def}

By the same way as above it is possible to obtain results similar those are formulated in Theorems \ref{T1}--\ref{fgsol}. In this case another Riemann surface with the same branch points arises but with a different basis of cycles, that corresponds to a different choice of jump matrices (Details can be found in~\cite{KS} for~$\Psi$ associated with the nonlinear Schr\"odinger equation).

The paper presents the matrix Baker--Akhiezer function associated with the Maxwell--Bloch equations. We used the matrix Riemann--Hilbert problem posed on the complex plane with a finite set of cuts. Such a Baker--Akhiezer function having the unit determinant, satisfies the AKNS equations for the Maxwell--Bloch system and generates the finite-gap quasi-periodic solution to the MB equations.

The matrix Baker--Akhiezer function will be useful for applying to Cauchy problems with periodic (quasi-periodic) finite-gap initial data as well as for the initial-boundary value problems with such type of initial and boundary functions. The suggested RH problem will be also useful for studying the long time/large space asymptotic behavior of solutions of different initial-boundary value problems to the MB equations by the way as, for example, in \cite{BT16,BK07a,BIK09,BK07,BKS11,BV07,BM13, BM14,D,DIZ93, DIZ97,DKMVZ99-1,DKMVZ99-2, DVZ98,DZ93, EGKT13,KMM03,KSZ15,KT12,KM10,KM12,MK06, MK10,TVZ04,TVZ06}. The focusing nonlinear Schr\"odinger equation and its finite-gap solutions are widely used for modeling of the so-called rogue waves. Some recent results in this field can be found in \cite{BET16,BG15,MS16}. In this regard, we hope that the results of paper will be useful for an investigation of the rogue waves of the Maxwell--Bloch equations.

\subsection*{Acknowledgments} The author thanks to the referees for careful reading of the manuscript and valuable recommendations.

\pdfbookmark[1]{References}{ref}
\LastPageEnding

\end{document}